\documentclass[12pt]{article}

\newif\ifplots 
\plotstrue 

\usepackage{graphicx}
\usepackage{newtxtext}
\usepackage{natbib}
\usepackage{hyperref}
\hypersetup{
    colorlinks = true,
    urlcolor   = blue,
    citecolor  = black,
}

\newcommand{\RomanNumeralCaps}[1]
\linenumbers

\usepackage{amsmath,amssymb,esint,bbm,textgreek,mathrsfs,mathtools}
\usepackage{setspace}
\usepackage{etoolbox}

\usepackage{stackengine}
\usepackage{scalerel}

\usepackage{cleveref}
\usepackage{algpseudocode}
\usepackage{algorithm}
\usepackage{caption}
\usepackage{subfigure}
\usepackage[normalem]{ulem}
\usepackage[table]{xcolor}
\usepackage{booktabs}
\usepackage{placeins}
\usepackage{diagbox}
\usepackage{tabularray}
\usepackage{makecell}
\usepackage{pdflscape}

\crefname{equation}{}{}

\graphicspath{{Figs/}}

\DeclareFontFamily{U}{mathx}{\hyphenchar\font45}
\DeclareFontShape{U}{mathx}{m}{n}{
      <5> <6> <7> <8> <9> <10>
      <10.95> <12> <14.4> <17.28> <20.74> <24.88>
      mathx10
      }{}
\DeclareSymbolFont{mathx}{U}{mathx}{m}{n}
\DeclareFontSubstitution{U}{mathx}{m}{n}
\DeclareMathAccent{\widecheck}{0}{mathx}{"71}
\makeatletter
\def\widebreve{\mathpalette\wide@breve}
\def\wide@breve#1#2{\sbox\z@{$#1#2$}%
     \mathop{\vbox{\m@th\ialign{##\crcr
\kern0.08em\brevefill#1{0.8\wd\z@}\crcr\noalign{\nointerlineskip}%
                    $\hss#1#2\hss$\crcr}}}\limits}
\def\brevefill#1#2{$\m@th\sbox\tw@{$#1($}%
  \hss\resizebox{#2}{\wd\tw@}{\rotatebox[origin=c]{90}{\upshape(}}\hss$}
\makeatletter
\newcommand{\A}{\mathcal{A}}

\newcommand{\E}{\mathcal{E}}

\newcommand{\R}{\mathcal{R}}

\newcommand{\OO}{\mathcal{O}}

\renewcommand{\S}{\mathcal{S}}

\newcommand{\RR}{\mathbb{R}}
\newcommand{\NN}{\mathbb{N}}

\newcommand{\grad}{\nabla}

\DeclareMathOperator*{\argmin}{arg\,min}
\newcommand{\Id}{\operatorname{Id}}

\newcommand\pp[2]{\frac{\partial^2 #1}{\partial #2^2}}

\newcommand{\Dpartial}[2]{ {\partial #1 \over \partial #2} }

\newcommand{\Dpartialn}[3]{ {\partial^{#3} #1 \over \partial #2^{#3}} }

\newcommand{\dt}{\Delta t}
\newcommand{\intO}{\int_{\Omega}}

\newcommand\STATE[1]{ s }
\newcommand\STATEin[1]{ s }

\newcommand{\Jl}{\mathcal{J_{\nu, \lambda}}}
\newcommand{\wphi}{\widehat{\phi}}

\def\Bmp#1{ \begin{minipage}{#1} }
\def\Emp{ \end{minipage} }
\def\Bmpc#1{ \begin{minipage}[c]{#1} }
\def\Bmpt#1{ \begin{minipage}[t]{#1} }
\def\Bmpb#1{ \begin{minipage}[b]{#1} }

\newcommand{\optphi}{\varphi_{\nu, \lambda}}

\renewcommand{\arraystretch}{1.5}
\definecolor{Gray}{gray}{0.9}
\newcolumntype{?}{!{\vrule width 2pt}}

\newcommand{\revt}[1]{{\color{black}#1}}
\newcommand{\revtt}[1]{{\color{black}#1}}

\newcommand\strike[2]{\bgroup\markoverwith
{\textcolor{red}{\rule[.5ex]{2pt}{0.4pt}}}\ULon{#1} \revtt{#2}}

\makeatletter
\renewcommand*\env@matrix[1][\arraystretch]{%
  \edef\arraystretch{#1}%
  \hskip -\arraycolsep
  \let\@ifnextchar\new@ifnextchar
  \array{*\c@MaxMatrixCols c}}
\makeatother

\usepackage{tikz,colortbl}
\usetikzlibrary{calc}
\usepackage{zref-savepos}
\newcounter{NoTableEntry}
\renewcommand*{\theNoTableEntry}{NTE-\the\value{NoTableEntry}}

\newcommand\p[2]{\frac{\partial #1}{\partial #2}}

\newtheorem{problem}{Problem}
\newtheorem{definition}{Definition}

\def\TT{{\mathbb{T}}}

\usepackage[T1]{fontenc}
\usepackage[utf8]{inputenc}
\usepackage{authblk}

\begin{document}
\title{Unraveling Self-Similar Energy Transfer Dynamics: a Case Study for 1D Burgers System}

\author[1,2]{Pritpal Matharu\thanks{Email address for correspondence: \texttt{matharu@mis.mpg.de}}}
\author[3]{Bartosz Protas}
\author[4]{Tsuyoshi Yoneda}
\affil[1]{Max Planck Institute for Mathematics in the Sciences, Leipzig,  Germany}
\affil[2]{Department of Mathematics, KTH Royal Institute of Technology, Stockholm, Sweden}
\affil[3]{Department of Mathematics and Statistics, McMaster University, Hamilton, Ontario, Canada}
\affil[4]{Graduate School of Economics, Hitotsubashi University, Tokyo, Japan}


\date{\today}
\maketitle

\begin{abstract}
  {In this work we consider the problem of \revt{constructing initial conditions for a flow model such that the resulting flow evolution leads to a} self-similar energy cascade consistent with Kolmogorov's statistical theory of turbulence. As a first step in this direction, we focus on the one-dimensional viscous Burgers equation as a toy model. Its solutions exhibiting self-similar behavior, in a precisely-defined sense, are found by framing this problems in terms of PDE-constrained optimization. The main physical parameters are the time window over which self-similar behavior is sought (equal to approximately one eddy turnover time), viscosity (inversely proportional to the ``Reynolds number") and an integer parameter characterizing the distance in the Fourier space over which self-similar interactions occur. Local solutions to this nonconvex PDE optimization problems are obtained with a state-of-the-art adjoint-based gradient method. Two distinct families of solutions, termed {\em viscous} and {\em inertial}, are identified and are distinguished primarily by the behavior of enstrophy which, {respectively,} uniformly decays and grows in the two cases. The physically meaningful and appropriately self-similar inertial solutions are found only when a sufficiently small viscosity is considered. These flows achieve the self-similar behaviour by a uniform steepening of the wave fronts present in the solutions. \revtt{The results obtained demonstrate that the proposed methodology may be used to search for self-similar behavior in more complex flow models, including shell models, 2D turbulence and, ultimately, 3D turbulence.}

}
\end{abstract}

\begin{flushleft}
\end{flushleft}




\section{Introduction\label{sec:intro}}

Self-similarity of the energy cascade is one of the most prominent statistical properties of hydrodynamic turbulence \citep{Frisch1995book}.
The energy cascade between a hierarchy of spatial and temporal scales was described {first} by Richardson \citep{Richardson1922book}, {who provided a} statistical account of this cascade in turbulent flows. This idea further leads to the concept {of} self-similar flow structures predicted by Kolmogorov \citep{Kolmogorov1941} who proposed a similarity hypothesis to explain the -5/3 power law describing the energy cascade. However, this approach is based on dimensional analysis of quantities defined in statistical terms and therefore is not directly related to the mathematical structure of the solutions of the Navier-Stokes system. In other words, being statistical in nature, this theory does not offer any insights into what kind of fluid motions can generate an energy cascade with the spectrum characterized by the -5/3 power law.

In 2017, Goto, Saito \& Kawahara \citep{Goto2017} used  direct numerical simulation (DNS) to reveal that the 3D Navier-Stokes turbulence in a periodic box has a hierarchical structure consisting of vortex tubes with a range of characteristic length scales. At each of these scales, the vortex tubes  generate a strain field that contributes to the generation of smaller-scale vortex tubes. Based on that result, Yoneda, Goto \& Tsurhashi \citep{Yoneda2022} reformulated Kolmogorov's -5/3 power law by defining an energy-transfer function in terms of vorticity \citep[equation (2.6)]{Yoneda2022} and then using it to identify a clear local-in-scale energy transfer structure from instantaneous spatially-averaged turbulence snapshots \citep[Figure 1]{Yoneda2022},\citep{Tsuruhashi2022}. However, that approach was based on spatial averaging and therefore it was not possible to identify the elementary processes, in the form of  specific {flow evolutions in time}, that give rise to a self-similar energy cascade. The purpose of this study is to address this question in the context of a simple toy model of turbulent flows, namely, the one-dimensional (1D) Burgers equation. A key novelty of our approach is that solutions with a certain self-similar structure are sought systematically by solving suitably-defined optimization problems for a partial differential equation (PDE). By focusing on a simplified model problem we aim to develop \revtt{and test an approach  that can be generalized to search for analogous self-similar flow phenomena in the context of more complex flow models, eventually including 3D turbulence.}

While PDE-constrained optimization has had a long history in fluid mechanics \citep{g03}, most of these studies have focused on solution of practical problems often motivated by engineering applications {\citep{Jameson1988,bmt01,Matharu2020,Matharu2022a}}. It was only recently that such techniques were employed to investigate fundamental questions concerning the structure and properties of solutions to hydrodynamic equations in a way that complements rigorous mathematical analysis \citep{p21a}. Problems tackled in this way include probing sharpness of a priori bounds on various energy-type quantities \citep{ap11a,ayala_doering_simon_2018}, nonlinear stability \citep{Kerswell2014}, dissipation anomaly \citep{MatharuProtasYoneda2022}, search for finite time singularities \citep{ap16,KangYunProtas2020,KangProtas2021,ZhaoProtas2023} and optimization of mixing \citep{HassanzadehChiniDoering2014,MilesDoering2018}. The present study builds on this line of research.

The main goal of this work is to find and analyze a family of \revtt{freely-decaying (unforced)} 1D Burgers flows featuring a self-similar energy cascade and parametrized by the value of the viscosity {coefficient} (or, equivalently, the Reynolds number) and a positive integer number characterizing the nonlocality of interactions in the Fourier space. The initial conditions for these flows are found as solutions of PDE-constrained optimization problems and are obtained numerically using a state-of-the-art adjoint-based optimization method. Our main finding is the evidence that families of such Burgers flows do indeed exist and that their self-similar behavior is realized by a uniform steepening of the fronts in the solutions. We also provide an analysis of how quantitative details of this process depend on physical parameters.

The structure of the paper is as follows: in the next section we introduce the model and state the optimization problem; in Section \ref{sec:approach} we describe the solution approach whereas the results are presented in Section \ref{sec:results}; discussion and final conclusions are deferred to Section \ref{sec:conclusions} while some technical material is collected in an appendix.

\section{The Governing Equation and Optimization Formulation}
\label{sec:model}

In this section we first introduce the 1D Burgers system and briefly discuss some properties of its solutions. It will be considered on the periodic domain $\Omega = \TT := [0,2\pi]$ and over the time window $[0,T]$, where the choice of the end time $T$ will be discussed below. Then, we formulate a PDE optimization problem designed to find solutions exhibiting a self-similar evolution.

\subsection{1D Burgers Equation\label{sec:burgers}}

The Burgers system in 1D has the following form subject to the periodic boundary conditions
\begin{subequations} \label{eq:Burgers}
\begin{alignat}{2} 
\p{u}{t} + u \p{u}{x} -\nu \pp{u}{x} &= 0 & \qquad &\text{in} \ (0, T] \times \Omega,  \label{eq:burg_eqn} \\
u(t=0) &= \phi & &\text{in} \ \Omega, \label{eq:BurgIC}
\end{alignat} 
\end{subequations}
where $\nu > 0$ is the viscosity and $\phi$ is the initial condition. It is known to admit globally defined smooth solutions for a broad class of initial data $\phi$ \citep{kl04}. \revt{There are three physical parameters in the problem, namely, the size $|\Omega|$ of the domain, the ``size" $U$ of the initial data $\phi$ and the viscosity coefficient $\nu$, which can be combined to form the dimensionless Reynolds number $Re = U |\Omega| / \nu$. In our study we will assume the first two parameters are fixed and will vary the viscosity coefficient.} Anticipating the formulation of the optimization problem below, we will assume here that $\phi \in H^1(\Omega)$, where $H^1(\Omega)$ is the Sobolev space of functions with square-integrable weak derivatives endowed with the inner product \citep{adams-2003}
\begin{equation}
\left\langle p_1, p_2 \right\rangle_{H^1(\Omega)} 
 {:=} \intO p_1  p_2 + \ell^2 \, \left(\p{p_1}{x} \, \p{p_2}{x} \right) \, dx,  \qquad \forall \  {p_1,p_2 \in H^1(\Omega)}, 
\label{eq:ipH1}
\end{equation}
where {``:='' means ``equal to by definition" and} $0 < \ell < \infty$ is an adjustable parameter (its significance will become evident in Section \ref{sec:approach}).

Various stochastic variants of system \eqref{eq:Burgers} have been used as models of turbulence \citep{bk07}, \revtt{typically in the forced setting realized by including a source term on the right-hand side (RHS) in \eqref{eq:burg_eqn}.} On a phenomenological level, equation \eqref{eq:burg_eqn} describes the competition between a nonlinear amplification in the form of wave steepening, which represents the forward energy cascade with energy transferred from large to small scales, and viscous dissipation \citep{FournierFrisch1983}. The wave steepening is characterized by the evolution of the enstrophy
\begin{equation}
    \E(u(t, \cdot)) := \int_0^{2\pi} \left| \Dpartial{u(t,x)}{x} \right|^2 \, dx =  \| u(t, \cdot) \|^2_{\dot{H}^1},
    \label{eq:E}
\end{equation}
which \revtt{in the unforced case} is well understood \citep{ap11a,p12,p12b}. In particular, sharp a priori bounds on the maximum growth of the enstrophy are now available \citep{AlbrittonDeNitti2023}. The triadic interactions responsible for the amplification of the enstrophy in 1D Burgers flows were analyzed in \citep{MurrayBustamante2018,buzzicotti2016phase,ProtasKangBustamante2024}.

\subsection{Optimization Problem\label{sec:optprob}}

Before we can state the optimization problem, we need to define {a suitable} notion of ``self-similar evolution". \revtt{While there is more than one way in which this property can be defined, we focus here on a formulation suitable for the freely-decaying problem we consider in which there is no statistical equilibrium.}
Let $\widehat{u}(t,k)$, $k \in \NN$, denote the Fourier coefficient of the solution $u(t,x)$ at some time $t \ge 0$. Given a positive integer $\lambda \in \NN^+$, we have the following
\begin{definition} \label{def:uself}
A solution $u = u(t,x)$ of \eqref{eq:Burgers} is said to be self-similar with an index $\lambda$ in terms of spectral energy transfer on $[t,t+T]$ if it satisfies the relation
\begin{equation}
 |\widehat{u}(t,k)|^2 = \lambda \, |\widehat{u}(t+T,\lambda k)|^2, \qquad k \in \NN, \quad t > 0.
 \label{eq:uself}
\end{equation}
\end{definition}
For brevity, we will refer to solutions satisfying this definition as simply ``self-similar". Intuitively, relation \eqref{eq:uself} can be interpreted to mean that during a self-similar evolution over the time $T$ a fraction $(1/\lambda) \in (0,1)$ of the kinetic energy \revt{initially contained in the large-scale motions} is transferred between the Fourier modes with wavenumbers $k$ and $\lambda k$. While existence of such solutions in a mathematically precise sense remains an open question, we will attempt here to construct them by finding initial conditions $\phi$ that minimize the functional $\Jl \; : \; H^1(\Omega) \rightarrow \RR$ defined as 
\begin{align} \label{eq:J}
\Jl(\phi) := \frac{1}{2} \, \sum_{k=0}^{\infty} {\underbrace{\left| |\widehat{\phi}(k)|^2 - \lambda |\widehat{u}(T, \lambda k; \phi)|^2 \right|}_{f(k;\phi)}}^2
\end{align} 
in which summation is restricted to positive wavenumbers $k$ only due to the conjugate symmetry $\widehat{u}(\cdot,k) = \overline{\widehat{u}(\cdot,-k)}$, where the overbar denotes complex conjugation, characterizing the Fourier coefficients of real-valued functions. \revtt{Notions of self-similarity other than given by Definition \ref{def:uself} may also be considered, especially in the context of forced flows where some form of statistical equilibrium holds. In such situations self-similarity can be defined by requiring, e.g., that energy spectra at different times collapse or that the energy flux exhibits some form of scale invariance.}

Clearly, in the absence of any constraints, the global minimum of $\Jl(\phi)$ is achieved by $\phi_0 \equiv 0$, since then $u(t;\phi_0) \equiv 0$, $t > 0$, and $\Jl(\phi_0) = 0$. In order to avoid such uninteresting trivial solutions, we need to constrain the initial condition to have a nonvanishing magnitude and we choose to do this by prescribing its enstrophy as $\E(\phi) = \E_0 > 0$. We thus have the following optimization problem
\begin{problem}\label{pb:1}
For fixed $T$, $\E_0, \nu \in \RR^+$ and $\lambda \in \mathbb{N}^+$ in \cref{eq:J}, find
\begin{equation*}
\optphi = \underset{\phi \in \S} {\argmin} \, \Jl(\phi), \quad \textrm{where} 
\quad \S := \left\{ \phi \in H^1(\Omega) \: : \: \int_0^{2\pi} \phi(x) \, dx = 0, \quad  \E(\phi) = \E_0 \right\}. 
\end{equation*}
\end{problem}
In the above $\S \subset H^1(\Omega)$ represents the constraint manifold which also includes the zero-mean condition (which is a property preserved in the evolution governed by \eqref{eq:burg_eqn}). The time window $T$ will be taken on the order of an eddy turnover time $t_e$ defined as 
\begin{align} \label{eq:te}
t_e := \sqrt{\frac{\nu}{\varepsilon}}, 
\end{align}
where $\varepsilon$ is the space-time averaged energy dissipation $\varepsilon :=  \frac{2\nu}{T |\Omega|} \int_{0}^{T}\E(u(t, \cdot))\,dt$ \revt{which is a well-defined quantity in freely-decaying Burgers flows. We add that the constraints defining the manifold $\S$ in Problem \ref{pb:1} are independent of the problem parameters and also remain well-defined when $\nu \rightarrow 0$.}


\section{Solution Approach\label{sec:approach}}

While methods of adjoint analysis are now well known \citep{g03}, we introduce our approach in some detail below given the somewhat nonstandard structure of the objective functional which is defined in the Fourier space \eqref{eq:J}. \Cref{pb:1} is Riemanian in the sense that its solutions are constrained to a Riemannian manifold $\S \subset H^1(\Omega)$ \citep{ams08}. It is also nonconvex and its local minimizers are approximated as $\optphi = \lim_{n\rightarrow \infty} \phi^{(n)}$ using a retracted discrete gradient flow
\begin{equation}
\begin{aligned}
\phi^{(n+1)} & =  \R_{\mathcal{S}}\left(\;\phi^{(n)} + \tau_n \nabla\Jl\left(\phi^{(n)}\right)\;\right), \qquad n = 1, 2, \dotsc, \\ 
\phi^{(1)} & =  \phi_1,
\end{aligned}
\label{eq:desc}
\end{equation}
where $\phi^{(n)}$ is an approximation of the minimizer obtained at the $n$-th iteration, $\phi_1 \in \S$ is the initial guess, $\R_{\mathcal{S}} \; : \; H^1(\Omega) \rightarrow \mathcal{S}$ is the retraction operator defined below  and $\tau_n$ is the length of the step in the direction of the gradient $\nabla\Jl(\phi^{(n)})$. To exploit the mathematical structure of \Cref{pb:1}, we will adopt the ``optimize-then-discretize" approach \citep{g03} where expressions for the different elements in \eqref{eq:desc} are derived in the continuous setting and discretized for the purpose of numerical evaluation at the final stage only; these steps are {documented} in Section \ref{sec:numerical} below. 

To obtain an expression for the gradient $\nabla\Jl(\phi)$ of the objective functional $\Jl(\phi)$, we begin by defining its G\^ateaux (directional) differential $\Jl'(\phi;\phi') := \lim_{\epsilon \rightarrow 0} \epsilon^{-1}\left[\Jl(\phi+\epsilon \phi') - \Jl(\phi)\right]$ for some arbitrary perturbation $\phi' \in H^1(\Omega)$ of the initial condition. Given the definition of the functional in \eqref{eq:J}, the G\^ateaux differential is
\begin{align}
\Jl'(\phi; \phi') & = \sum_{k=0}^{\infty} \left[ |\wphi(k)|^2 - \lambda \, |\widehat{u}(T, \lambda k; \phi)|^2 \right] \, \left[ \wphi'(k)\overline{\wphi}(k) + \wphi(k)\overline{\wphi}'(k) \right] \nonumber\\
& - \sum_{k=0}^{\infty} \lambda \left[ |\wphi(k)|^2 - \lambda \, |\widehat{u}(T, \lambda k; \phi)|^2 \right] \, \left[\widehat{u}'(T, \lambda k; \phi, \phi')\overline{\widehat{u}}(T, \lambda k; \phi)\right] \nonumber \\
&- \sum_{k=0}^{\infty} \lambda \left[ |\wphi(k)|^2 - \lambda \, |\widehat{u}(T, \lambda k; \phi)|^2 \right] \, \left[\widehat{u}(T, \lambda k; \phi)\overline{\widehat{u}}'(T, \lambda k; \phi, \phi') \right], \nonumber \\
& = 2\sum_{k=0}^{\infty} \left[ |\wphi(k)|^2 - \lambda \, |\widehat{u}(T, \lambda k; \phi)|^2 \right] \wphi(k)\,\overline{\wphi}'(k) \nonumber\\
& -\underbrace{2\sum_{k=0}^{\infty} \lambda \left[ |\wphi(k)|^2 - \lambda \, |\widehat{u}(T, \lambda k; \phi)|^2 \right]  \widehat{u}(t, \lambda k; \phi)\,\overline{\widehat{u}}'(t, \lambda k; \phi, \phi')}_{\A(\phi;\phi')}, \label{eq:dJl}
\end{align}
where we used the identity $\sum_{k=0}^{\infty} \, \widehat{\upsilon}_1(k)\overline{\widehat{\upsilon}}_2(k) = \sum_{k=0}^{\infty} \, \overline{\widehat{\upsilon}}_1(k){\widehat{\upsilon}_2}(k)$ valid for real-valued functions $v_1(x)$, $v_2(x)$ and $u'(t, x; \phi, \phi')$ satisfies the perturbation system 
\begin{subequations} \label{eq:Burgersp}
\begin{alignat}{2} 
\p{u'}{t} + u' \p{u}{x} + u \p{u'}{x} -\nu \pp{u'}{x} &= 0 & \qquad &\text{in} \ (0, T] \times \Omega,  \label{eq:burg_eqnp} \\
u'(t=0) &= \phi' & &\text{in} \ \Omega
\label{eq:BurgICp}
\end{alignat} 
\end{subequations}
obtained by linearizing system \eqref{eq:Burgers} around the solution $u(\cdot,\cdot;\phi)$. Noting that when viewed as a function of its second argument ($\phi'$) and for a fixed $\phi$, the G\^ateaux differential \eqref{eq:dJl} is a bounded linear functional on both $H^1(\Omega)$ and $L^2(\Omega)$, and we can invoke the Riesz representation theorem \citep{b77}
\begin{equation} 
\label{eq:Riesz}
{\Jl}'(\phi; \phi') = \left\langle \grad^{H^1}{\Jl}, {\phi}' \right\rangle_{H^1(\Omega)} = \left\langle \grad^{L^2}{\Jl}, {\phi}' \right\rangle_{L^2(\Omega)},
\end{equation}
where the inner product $\langle\cdot,\cdot\rangle_{L^2(\Omega)}$ is obtained from \eqref{eq:ipH1} by setting $\ell = 0$. It is convenient to first obtain an expression for the $L^2$ gradient $\grad^{L^2}{\Jl}$ and then use identity \eqref{eq:Riesz} together with \eqref{eq:ipH1} to deduce the required Sobolev gradient $\grad{\Jl} = \grad^{H^1}{\Jl}$ \citep{pbh04}. We note that the last term in expression \eqref{eq:dJl} for the G\^ateaux differential, $\A(\phi;\phi')$, is not consistent with the Riesz representation \eqref{eq:Riesz} since the arbitrary perturbation $\phi'$ does not appear in it explicitly as a linear factor and is instead ``hidden" as the initial condition in the perturbation system \eqref{eq:Burgersp}. To transform this term to the required form, we use the adjoint calculus. Multiplying \eqref{eq:burg_eqnp} by the adjoint state $u^* \; : \; [0, T] \times \Omega \rightarrow \RR$, integrating the resulting expression over $[0, T] \times \Omega$ and performing integration by parts with respect to $t$ and $x$, we obtain
\begin{align}
0 &= \int_0^T \int_0^{2\pi} \left(\p{u'}{t} + u' \p{u}{x} + u \p{u'}{x} -\nu \pp{u'}{x} \right) u^*  \, dx \, dt \nonumber \\
& = \int_0^{2\pi} \Bigg[ u^*(T, x)u'(T, x) - u^*(0, x)u'(0, x) + \int_0^T \left( -\p{u^{*}}{t} - u \p{u^{*}}{x}-\nu \pp{u^{*}}{x} \right) {u}' \, dt \Bigg] \, dx, \nonumber\\
\label{eq:dual_BurgAdj}
\end{align}
where all the boundary terms resulting from integration by parts in space vanish due to the periodic boundary conditions. Defining the adjoint system as
\begin{subequations} \label{eq:BurgAdj}
\begin{alignat}{2} 
-\p{u^{*}}{t} - u \p{u^{*}}{x} - \nu \pp{u^{*}}{x} &= 0 &\qquad &\text{in} \ (0, T] \times \Omega,  \label{eq:adjeqn} \\
u^{*}(t=T) &= W(\phi) & &\text{in} \ \Omega, \label{eq:adjIC}
\end{alignat} 
\end{subequations}
with the terminal condition stated in Fourier space as
\begin{align}\label{eq:W}
\widehat{W}(k;\phi) = 
\begin{cases}
-\frac{2\lambda}{\pi} \left[ |\wphi(k/\lambda)|^2 - \lambda \, |\widehat{u}(T, k; \phi)|^2 \right] \widehat{u}(T, k; \phi), \quad &k = \lambda \, n, \qquad n \in \mathbb{N}\\
\quad 0, &\text{otherwise}
\end{cases}
\end{align}
reduces \eqref{eq:dual_BurgAdj} to
\begin{equation}
 \A(\phi;\phi') = \int_0^{2\pi} u^*(T, x)u'(T, x) \, dx =\int_0^{2\pi} u^*(0, x)\phi'(x) \, dx = \Big\langle u^*(0, \cdot),
\phi' \Big\rangle_{L^2(\Omega)},
\label{eq:dual2}
\end{equation}
where Parseval's theorem was used to express integrals of products of functions in terms of sums of products of their Fourier coefficients. The adjoint system \eqref{eq:BurgAdj} is a terminal-value problem with the terminal condition \eqref{eq:W} determined by the form {of} the G\^ateaux differential \eqref{eq:dJl}, cf.~\eqref{eq:dual2}, and hence also by the objective functional \eqref{eq:J}. Therefore, the system needs to be integrated backwards in time and its coefficients $u(t,x;\phi)$ are given by the solution of the governing system \eqref{eq:Burgers} around which linearization is performed.

Combining \eqref{eq:dJl} and \eqref{eq:dual2}, we obtain the following expression for the $L^2$ gradient of the functional \eqref{eq:J} given in terms of its Fourier-space representation as
\begin{align} \label{eq:gradL2}
{\widehat{\grad^{L^2}{\Jl}}(k)} = \frac{2}{\pi} \, \left[ |\wphi(k)|^2 - \lambda \, |\widehat{u}(T, \lambda k; \phi)|^2 \right] \wphi(k) + \widehat{u}^{*}(0, k), \qquad k \in \NN.
\end{align}

Since the $L^2$ gradient does not possess the regularity needed to construct solutions of  \Cref{pb:1}, we must determine the corresponding Sobolev $H^1$ gradient. 
Using the second equality in the Riesz identity \eqref{eq:Riesz} together with the definition \eqref{eq:ipH1} of the $H^1$ inner product, performing integration by parts and noting the arbitrariness of $\phi'$, we obtain the required $H^1$ gradient {$\grad{\Jl}$} as a solution of the elliptic boundary-value problem
\begin{equation}
\left[ \Id \, - \,\ell^2 \,\Dpartialn{}{x}{2} \right] {\grad{\Jl}
= \grad^{L^2}{\Jl}}  \qquad \text{in} \ \Omega, 
\label{eq:gradH1}
\end{equation}
subject to the periodic boundary conditions.  As shown in \citep{pbh04}, extraction of gradients in spaces of smoother functions such as $H^1(\Omega)$ can be interpreted as low-pass filtering of the $L^2$ gradients with {the} parameter $\ell$ acting as the cut-off length-scale. 

The retraction operator in \eqref{eq:desc} is defined in terms of the normalization 
\begin{equation} 
\R_{\mathcal{S}}(\phi)  {:=} \sqrt{\frac{\E_0}{\E(\phi)}}\,\phi .
\label{eqAnomalDiss:R}
\end{equation}
An optimal step size $\tau_n$ can be determined by solving the
minimization problem
\begin{equation}
\tau_n = \underset{\tau > 0}{\argmin}\left\{ \Jl\left( \R_{\mathcal{S}} \left( \phi^{(n)} + \tau \, \grad\Jl(\phi^{(n)}) \right) \right) \right\},
\label{eqAnomalDiss:tau_n}
\end{equation}
which can be interpreted as a modification of a standard line search problem \citep{nw00} with optimization performed following an arc (a geodesic) lying on the constraint manifold $\mathcal{S}$, rather than a straight line. A numerical implementation of the approach described here is presented below.

\subsection{Numerical Implementation \label{sec:numerical}}
The governing system and the corresponding adjoint system, respectively \cref{eq:Burgers} and \cref{eq:BurgAdj}, are discretized in space using a Fourier pseudo-spectral method. Nonlinear products are evaluated in physical space and appropriately dealiased using the 2/3 rule combined {with} a Gaussian filter introduced in \citep{hou2009blow}. Time integration is performed using a four-step, implicit/explicit Runge-Kutta scheme which yields globally {third-order} accuracy \citep{Alimo2020}. Since the number of spatial discretization points $N$ and temporal step-size $\dt$ heavily depend on the selected values of $\nu$ and $\lambda$, we defer the discussion of selection of these parameters to the next section. Nonetheless, validation of the gradient \cref{eq:gradL2}, which is a key element of the optimization approach and requires the numerical solution of both {systems} \cref{eq:Burgers} and \cref{eq:BurgAdj}, is given in the \Cref{sec:kappa}.

The Sobolev gradient is determined using $\ell = 10$ in \eqref{eq:ipH1} and the corresponding elliptic boundary-value problem \cref{eq:gradH1} is solved using the {Fourier-Galerkin} spectral method. The line minimization problem \cref{eqAnomalDiss:tau_n} is solved using Brent's algorithm \citep{press2007numerical}. Iterations \eqref{eq:desc} were declared converged when a stopping criterion given in terms of the relative decrease of the cost functional, $|\Jl(\phi^{(n)}) - \Jl(\phi^{(n-1)})|/\Jl(\phi^{(n-1)}) < 1 \times 10^{-8}$, 
was satisfied or the number of iteration exceeded the limit of $n=1,000$. To accelerate convergence of the iterations in \eqref{eq:desc}, the gradient $\nabla\Jl\left(\phi^{(n)}\right)$ in \cref{eq:desc} was replaced with a conjugate gradient constructed using the Polak-Ribi\`ere formula \citep{nw00}. 
Our approach is implemented in MATLAB and the code is available {in} {\citep{MatharuGitBurgersEnergyTransfer}}. 

\section{Results\label{sec:results}}

In this section we present results obtained by solving \Cref{pb:1} with $\E_0 = \pi$ and with $\lambda$ and $\nu$ varying over a broad range of values, namely, $\lambda = 2, \dots, 7$ and $\nu = 2.5 \times 10^{-3}, 2.5 \times 10^{-4}, 2.5 \times 10^{-5}, 2.5 \times 10^{-6}, 2.5 \times 10^{-7}$. \revt{We recall that, with the initial enstrophy $\E_0$ and the domain $\Omega$ both fixed, the characteristic ``size" of the initial data can be expressed as $U = \sqrt{|\Omega| \E_0}$ and} decreasing the viscosity coefficient $\nu$ can be interpreted as increasing the ``Reynolds number" of the flow. 
When $\lambda$ increases and $\nu$ decreases, the numerical resolution both in space and in time needs to be refined in order for the solutions $u(t,x;\phi)$ to be well resolved, which can be checked by examining their Fourier spectra. The values of $N$ and $\dt$ used for different $\lambda$ and $\nu$ are summarized in \Cref{tbl:param}. As regards the choice of the time window $T$, our goal is for it to be on the order of an eddy turnover time \eqref{eq:te}, i.e., $T = C_T \, t_e$, where $C_T = \OO(1)$. However, $t_e$ depends on the solution $u(t,x;\optphi)$ of \Cref{pb:1} and therefore the actual value of the constant $C_T$ can only be determined a posteriori, i.e., after \Cref{pb:1} has been solved. Hence, in our computations we use a fixed time window $T = 0.6$ and the resulting values of the constant $C_T$ are reported in \Cref{tbl:te}. We have experimented with many other time windows $T$, both shorter and longer, and the results obtained were qualitatively similar to those reported below.

In order to illustrate the performance of algorithm \eqref{eq:desc}, in \Cref{fig:Jdecrease}a we show the dependence of the objective functional $\Jl(\phi^{(n)})$ on iterations $n$ for the solutions of 
\Cref{pb:1} obtained with $\nu =2.5 \times 10^{-5}$ and $\lambda = 2,4,6$. We see that the values of the objective functional drop significantly, down to $\OO(10^{-8}-10^{-9})$ with smaller values achieved for smaller $\lambda$. \revt{To verify that the flows $u(t,x;\optphi)$ do indeed satisfy the definition of self-similarity \cref{eq:uself} with a good accuracy,} expressions $f(k,\optphi)$, cf.~\eqref{eq:J}, corresponding to the optimal solutions $\optphi$ obtained for different $\lambda$ are shown as functions of $k$ in Figure~\ref{fig:Jdecrease}b. We see that in all cases these expressions are \revt{quite small, not exceeding $\mathcal{O}(10^{-5})$, and are} generally decreasing functions of the wavenumber $k$.

\revt{To verify that the optimal flows we found are not numerical artefacts resulting from a finite numerical resolution, we solve Problem \ref{pb:1} with $\nu = 2.5 \times 10^{-5}$ and $\lambda = 2, 4, 6$ again, but with a refined numerical resolution. More specifically, the number of grid points $N$  is increased by a factor of two as compared to the values indicated in \Cref{tbl:param} and the time step fixed to $\Delta t = 2 \times10^{-4}$. In all cases, the solution $\optphi$ obtained by solving Problem \ref{pb:1} with the original numerical parameters was used as the initial guess $\phi_{1}$ in \eqref{eq:desc}. The dependence of the objective functional $\Jl(\phi^{(n)})$ normalized with respect to $\Jl(\phi_{1})$ on iterations $n$  in these calculations is shown in Figure~\ref{fig:Jrefine}. We see that this resolution refinement produces only a minuscule (less than $15\%$) further reduction of the objective functional. This should be contrasted with the decrease of the objective functional, typically involving several orders of magnitude (cf.~\Cref{fig:Jdecrease}a), which occurs when Problem \ref{pb:1} is solved ``from scratch" with a fixed numerical resolution. These observations thus allow us to conclude that solutions of Problem \ref{pb:1} obtained with refined numerical resolutions produce convergent approximations of well-defined optimizers $\optphi$. Moreover, since the obtained values of the objective functional are very small and approach zero (i.e., the global minimum) as the numerical resolution is refined, this suggests that these optimizers are in fact global.}

Due to the nonlinearity of the governing system \eqref{eq:Burgers}, \Cref{pb:1} is nonconvex and as such may admit multiple local minimizers. This is indeed the situation we encounter at least for some combinations of $\nu$  and $\lambda$. More specifically, we have found evidence for the existence of two distinct classes of solutions of \Cref{pb:1} that we refer to as {\it viscous} and {\it inertial}. Their main features are contrasted in \Cref{fig:ViscInComp} and are summarized below.
\begin{itemize}
    \item {\bf Viscous solutions} involve highly oscillatory optimal initial conditions $\optphi$, cf.~Figure~\ref{fig:ViscInComp}a, with energy concentrated at high wavenumbers $k$ in the dissipative subrange, cf.~Figure~\ref{fig:ViscInComp}e. Consequently, the resulting evolution does not involve any inertial transfer of energy between different Fourier modes and instead the solutions are {quickly} attenuated by the viscous dissipation, cf.~Figure~\ref{fig:ViscInComp}c (which is the origin of the name we gave these solutions). The enstrophy $\E(u(t,\cdot;\optphi))$ of viscous solutions is a rapidly decreasing function of time $t$, cf.~Figure \ref{fig:ViscInComp}h. As they do not feature any large-scale structure and have a small magnitude, the viscous solutions resemble the minimizer $\phi_0 \equiv 0$ of functional \eqref{eq:J} in the absence the constraint $\E(\phi) = \E_0 >0$ (see the discussion before the statement of \Cref{pb:1}).

    \item {\bf Inertial solutions} minimize the objective functional \eqref{eq:J} via inertial transfer of energy to Fourier modes with higher wavenumbers $k$, cf.~Figure~\ref{fig:ViscInComp}f, which in the physical space is achieved by a self-similar steepening of the {wave} fronts, cf.~Figures~\ref{fig:ViscInComp}b,d. As a result, the enstrophy $\E(u(t,\cdot;\optphi))$ of inertial solutions is a rapidly increasing function of time $t$, cf.~Figure \ref{fig:ViscInComp}h. In contrast to the viscous solutions, the magnitude of the inertial solutions does not change much over the time window $T$.
\end{itemize}
{Viscous solutions are ``generic"} as they can be found by solving \Cref{pb:1} using essentially any initial guess $\phi_1$. On the other hand, inertial solutions are ``rare", in the sense that they can only be found if a ``good" initial guess $\phi_1$ is used in \eqref{eq:desc}. The final values of the objective functional \eqref{eq:J} are in both cases comparable, cf.~Figure~\ref{fig:Jdecrease}a, although the expressions $f(k,\optphi)$ tend to have a different dependence on $k$ for the viscous and inertial solutions, cf.~Figure~\ref{fig:ViscInComp}g. Moreover, for a given value of $\lambda$ the inertial solutions can only be found provided the viscosity coefficient $\nu$ is sufficiently small. This is because the inertial solutions involve a self-similar energy transfer between Fourier modes with wavenumbers $k$ and $\lambda k$, which can only happen if the inertial range is sufficiently wide or, {equivalently, if} the Reynolds number is sufficiently high. A key qualitative feature differentiating viscous and inertial solutions is the time evolution of the enstrophy \eqref{eq:E} which is, respectively, a decreasing and {an} increasing function of time for these two families of solutions, cf.~Figure \ref{fig:ViscInComp}h. \Cref{tbl:te} provides information about the values of $\lambda$ for which inertial solutions could be found for decreasing values of $\nu$, {and} also includes the number of wave fronts, $P$, present in the solution at the final time (during the evolution from $t=0$ to $t = T$ this number typically decreases by no more than 15\%). Needless to say, only the inertial solutions are self-similar in the sense of \Cref{def:uself} and hence we will exclusively focus on them in the remainder of this work.

\revtt{The observation that inertial solutions are ``rare", in the sense discussed above, raises the question about their robustness understood as persistence of self-similar evolution under perturbations of the optimal initial conditions $\optphi$. To address this question, we solve the governing system \eqref{eq:Burgers} over the time interval $[0,T]$ for a family of initial data $\phi^{\eta}$ obtained by adding random noise of magnitude $\eta \in \RR^{+}$ to the optimal initial condition $\optphi$ and then evaluating the objective functional \eqref{eq:J} as function of $\eta$. The perturbed initial condition is obtained by adding random perturbations to the Fourier coefficients of $\optphi$, i.e., as 
\begin{equation}
\phi^{\eta}(x) = \sum_{k = -N/2+1}^{N/2} \left[\widehat{\optphi}\right]_{k} (1+\eta \theta_{k}) e^{i k x},
\label{eq:phieta}	
\end{equation}
where $\theta_{k}$ are independent samples of a normally-distributed random variable with zero mean and unit variance, and then normalizing $\phi^{\eta}$ such that the constraint $\E(\phi^{\eta}) = \E_{0}$ is satisfied. The results obtained for $\nu = 2.5 \times 10^{-5}$ and $\lambda = 2, 4, 6$ are shown in Figure~\ref{fig:Jpert} where we see that flow evolutions cease to be self-similar, which is marked by the increase of $\Jl(\phi^{\eta})$ above its baseline value corresponding to $\eta = 0$, only provided the perturbations are sufficiently large, i.e., $\eta \gtrapprox 10^{-4}$ or $\eta \gtrapprox 10^{-2}$ depending on $\lambda$. For values of $\eta$ larger than this threshold, we have $\Jl(\phi^{\eta}) = \OO(\eta^{2})$ which, since the functional \eqref{eq:J} is quartic in $\phi$, suggests that the departure from self-similarity is only sublinear in $\eta$. These findings indicate that while inertial solutions are ``rare", they are in fact relatively robust.}

\revt{Finding optimal Burgers flows $u(t,\cdot;\optphi)$ as solutions of \Cref{pb:1} ensures these flows are self-similar on the time interval $[0,T]$ and a natural question is what happens with this property at later times $t > T$. To answer this question, we consider the function $g(t;\optphi) := \frac{1}{2} \, \sum_{k=0}^{\infty} \left| |\widehat{u}(t, \lambda k; \optphi)|^2 - \lambda |\widehat{u}(t+T, \lambda k; \optphi)|^2 \right|$ which is the expression from the objective functional \eqref{eq:J} evaluated over a sliding window $[t,t+T]$.  We plot this function for $t \in [T,2T]$ in Figure \ref{fig:JSlide} for the optimal Burgers flows obtained for $\nu = 2.5 \times 10^{-5}$ and $\lambda = 2, 4, 6$. Unsurprisingly, the property of self-similarity is quickly lost. However, it is interesting to observe that the loss of self-similarity is much slower for larger values of $\lambda$.
}

We now examine how properties of the inertial solutions vary in function of the parameters $\nu$ and $\lambda$. Families of optimal initial conditions $\optphi$ were obtained using a continuation approach where \Cref{pb:1} was solved for some $\nu_1$ and $\lambda_1$ using either $\varphi_{\nu_2,\lambda_1}$ or $\varphi_{\nu_1,\lambda_2}$ for some $\RR^+ \ni \nu_2 > \nu_1$ and $\NN^+ \ni \lambda_2 < \lambda_1$ as the initial guess $\phi_1$ in \eqref{eq:desc}. The solutions found for fixed viscosity coefficients $\nu$ and increasing values of $\lambda$ are summarized in \Cref{tbl:LambComp}, where we see that when $\lambda > 2$ inertial solutions can be found only provided $\nu$ is sufficiently small. The reason for this is clear when we examine the Fourier spectra of the optimal initial conditions $\optphi$ and of the corresponding final states $u(T,\cdot;\optphi)$ (the third and fourth columns in \Cref{tbl:LambComp}) --- we see that during its evolution the solution develops a well-defined inertial range where energy is cascaded in a self-similar manner minimizing the objective functional \eqref{eq:J}. When the viscosity is large, this inertial range does not persist long enough to accommodate energy transfer between Fourier coefficients with a large separation in the wavenumber space, which needs to happen when $\lambda$ is large.

In terms of the physical-space representation, the self-similar energy cascade is realized by the steepening of the wave fronts which is clearly visible in the first and second columns of \Cref{tbl:LambComp}. In other words, during the evolution over $[0,T]$ the global structure of the solutions remains essentially unchanged, but all the wave fronts steepen in almost the same manner. \revt{For beyond the time window $T$ the steepened wave fronts form shocks, for which afterwards diffusion dominates and dissipates the flow.} There are some apparent similarities in how this process unfolds in solutions obtained with even and odd values of $\lambda$. When $\nu$ is fixed, the number of distinct fronts in the solution is an increasing function of $\lambda$ which is illustrated in \Cref{fig:dudxT}. \Cref{tbl:te} provides more details about this trend, although additional data would be needed in order to determine the precise form of the dependence of $P$ on $\lambda$ for a fixed $\nu$.
Finally, in \Cref{tbl:ViscComp} we characterize solutions obtained by solving \Cref{pb:1} with $\lambda = 2,5$ and different values of $\nu$ (this is a subset of the results already shown in \Cref{tbl:LambComp}, but organized in a different manner). We see that, interestingly, both the optimal initial conditions $\optphi$ and the corresponding final states $u(T,x;\optphi)$ appear to converge to well-defined limits as $\nu \rightarrow 0$ while their Fourier spectra become more developed.

~\\
\begin{landscape}

\begin{table}[htbp]
\centering
\begin{tblr}{
  colspec={ |c|c|c|c|c|c|c|},
  hline{1,2,4,6,8,10,12}, hline{2} =2\arrayrulewidth, vline{2} = 2\arrayrulewidth,
  cell{2}{1} = {r=2}{m}, cell{4}{1} = {r=2}{m}, cell{6}{1} = {r=2}{m}, cell{8}{1} = {r=2}{m}, cell{10}{1} = {r=2}{m}, 
  cell{2}{6} = {r=2}{m}, cell{2}{7} = {r=2}{m}, cell{8}{3} = {r=2}{m}, cell{8}{4} = {r=2}{m}, cell{8}{5} = {r=2}{m}, cell{8}{6} = {r=2}{m}, cell{8}{7} = {r=2}{m}, cell{10}{3} = {r=2}{m}, cell{10}{4} = {r=2}{m}, cell{10}{5} = {r=2}{m}, cell{10}{6} = {r=2}{m}, cell{10}{7} = {r=2}{m} 
}
\backslashbox[2.5cm]{\hspace{1.0cm}$\nu$}{$\lambda$} & $2$ & $3$ & $4$ & $5$ & $6$ & $7$\\
$2.5 \times 10^{-3}$ & $N=4096$ & \SetCell[c=1]{c, red!15}$N=8192$ & \SetCell[c=1]{c, red!15}$N=8192$ & \SetCell[c=1]{c, red!15}$N=16384$ & \SetCell[c=1]{c, black!60} & \SetCell[c=1]{c, black!60} \\ 
 & {\bf $\dt = 1 \times 10^{-3}$} & \SetCell[c=1]{c, red!15}{\bf $\dt = 1 \times 10^{-3}$} & \SetCell[c=1]{c, red!15}{\bf $\dt = 1 \times 10^{-3}$} & \SetCell[c=1]{c, red!15}{\bf $\dt = 1 \times 10^{-3}$} &  & \\ 
$2.5 \times 10^{-4}$ & $N=8192$ & $N=16384$ & $N=16384$ & $N=32768$ & \SetCell[c=1]{c, red!15}$N=65536$ & \SetCell[c=1]{c, red!15}$N=65536$\\ 
 & {\bf $\dt = 5 \times 10^{-4}$} & {\bf $\dt = 5 \times 10^{-4}$} & {\bf $\dt = 5 \times 10^{-4}$} & {\bf $\dt = 5 \times 10^{-4}$} & \SetCell[c=1]{c, red!15}{\bf $\dt = 5 \times 10^{-4}$} & \SetCell[c=1]{c, red!15}{\bf $\dt = 5 \times 10^{-4}$} \\ 
$2.5 \times 10^{-5}$ & $N=16384$ & { N=131072} & { N=131072} & { N=131072} & { N=131072} & { N=131072} \\ 
 & {\bf $\dt = 5 \times 10^{-4}$} & {\bf $\dt = 2 \times 10^{-4}$} & {\bf $\dt = 2 \times 10^{-4}$} & {\bf $\dt = 2 \times 10^{-4}$} & {\bf $\dt = 2 \times 10^{-4}$} & {\bf $\dt = 2 \times 10^{-4}$} \\ 
$2.5 \times 10^{-6}$ & $N=16384$ & \SetCell[c=1]{c, black!60} & \SetCell[c=1]{c, black!60} & \SetCell[c=1]{c, black!60} & \SetCell[c=1]{c, black!60} & \SetCell[c=1]{c, black!60}\\ 
 & {\bf $\dt = 5 \times 10^{-4}$} &  &  &  &  & \\ 
$2.5 \times 10^{-7}$ & $N=16384$ & \SetCell[c=1]{c, black!60} & \SetCell[c=1]{c, black!60} & \SetCell[c=1]{c, black!60} & \SetCell[c=1]{c, black!60} & \SetCell[c=1]{c, black!60} \\ 
 & {\bf $\dt = 5 \times 10^{-4}$} &  &  &  &  &  
\end{tblr}
\caption{The number of spatial discretization points $N$ and the temporal step size $\dt$ used to solve \Cref{pb:1} for different values of the parameters $\nu$ and $\lambda$. Pink background in a cell indicates that no inertial solutions were found for the given values of $\nu$ and $\lambda$. If a cell is left blank, then no attempt was made to solve \Cref{pb:1} with the corresponding values of $\nu$ and $\lambda$.}
\label{tbl:param}
\end{table}
\end{landscape}

\begin{landscape}
\begin{table}[htbp]
\centering
\begin{tblr}{
  colspec={ |c|c|c|c|c|c|c|},
  hline{1,2,4,6,8,10,12}, hline{2} =2\arrayrulewidth, vline{2} = 2\arrayrulewidth,
  cell{2}{1} = {r=2}{m}, cell{4}{1} = {r=2}{m}, cell{6}{1} = {r=2}{m}, cell{8}{1} = {r=2}{m}, cell{10}{1} = {r=2}{m}, 
  cell{2}{6} = {r=2}{m}, cell{2}{7} = {r=2}{m}, cell{8}{3} = {r=2}{m}, cell{8}{4} = {r=2}{m}, cell{8}{5} = {r=2}{m}, cell{8}{6} = {r=2}{m}, cell{8}{7} = {r=2}{m}, cell{10}{3} = {r=2}{m}, cell{10}{4} = {r=2}{m}, cell{10}{5} = {r=2}{m}, cell{10}{6} = {r=2}{m}, cell{10}{7} = {r=2}{m},  
  cell{2}{3} = {r=2}{m}, cell{2}{4} = {r=2}{m}, cell{2}{5} = {r=2}{m}, cell{4}{6} = {r=2}{m}, cell{4}{7} = {r=2}{m} 
}
\backslashbox[2.5cm]{\hspace{1.0cm}$\nu$}{$\lambda$} & $2$ & $3$ & $4$ & $5$ & $6$ & $7$\\
$2.5 \times 10^{-3}$ & $C_T = 0.70$ & \SetCell[c=1]{c, red!15} $C_T = 0.09$ & \SetCell[c=1]{c, red!15}$C_T = 0.09$ & \SetCell[c=1]{c, red!15}$C_T = 0.21$ & \SetCell[c=1]{c, black!60} & \SetCell[c=1]{c, black!60} \\ 
 & {$P = 9\phantom{0}$} & \SetCell[c=1]{c, red!15}{~} & \SetCell[c=1]{c, red!15}{~} & \SetCell[c=1]{c, red!15}{~} &  & \\ 
$2.5 \times 10^{-4}$ & $C_T = 0.75$ & $C_T = 0.81$ & $C_T = 0.85$ & $C_T = 0.72$ & \SetCell[c=1]{c, red!15}$C_T = 0.51$ & \SetCell[c=1]{c, red!15}$C_T = 0.53$\\ 
 & {$P = 10$} & {$P = 19$} & {$P = 15$} & {$P = 27$} & \SetCell[c=1]{c, red!15}{~} & \SetCell[c=1]{c, red!15}{~} \\ 
$2.5 \times 10^{-5}$ & $C_T = 0.75$ & $C_T = 1.44$ & $C_T = 0.90$ & $C_T = 1.35$ & $C_T = 0.67$ & $C_T = 0.86$ \\ 
 & {$P = 10$} & {$P = 19$} & {$P = 20$} & {$P = 28$} & {$P = 40$} & {$P = 48$} \\ 

$2.5 \times 10^{-6}$ & $C_T = 0.74$ & \SetCell[c=1]{c, black!60} & \SetCell[c=1]{c, black!60} & \SetCell[c=1]{c, black!60} & \SetCell[c=1]{c, black!60} & \SetCell[c=1]{c, black!60}\\ 
 & {$P = 10$} &  &  &  &  & \\ 
$2.5 \times 10^{-7}$ & $C_T = 0.74$ & \SetCell[c=1]{c, black!60} & \SetCell[c=1]{c, black!60} & \SetCell[c=1]{c, black!60} & \SetCell[c=1]{c, black!60} & \SetCell[c=1]{c, black!60} \\ 
 & {$P = 10$} &  &  &  &  & \\ 
\end{tblr}
\caption{Values of the constant $C_T = T / t_e$ relating the optimization time window $T$ and the eddy turnover time $t_e$, cf.~\eqref{eq:te}, for different values of $\nu$ and $\lambda$. In addition, for inertial solutions, we show the number of wave fronts $P$ present in the solution at final time. Pink background in a cell indicates that no inertial solutions where found for the given values of $\nu$ and $\lambda$ (and hence $P$ was not determined). If a cell is left blank, then no attempt was made to solve \Cref{pb:1} with the corresponding values of $\nu$ and $\lambda$.}
\label{tbl:te}
\end{table}
\end{landscape}

\ifplots{
\begin{figure}\centering
  \subfigure[]
  {
    \includegraphics[scale=0.4]{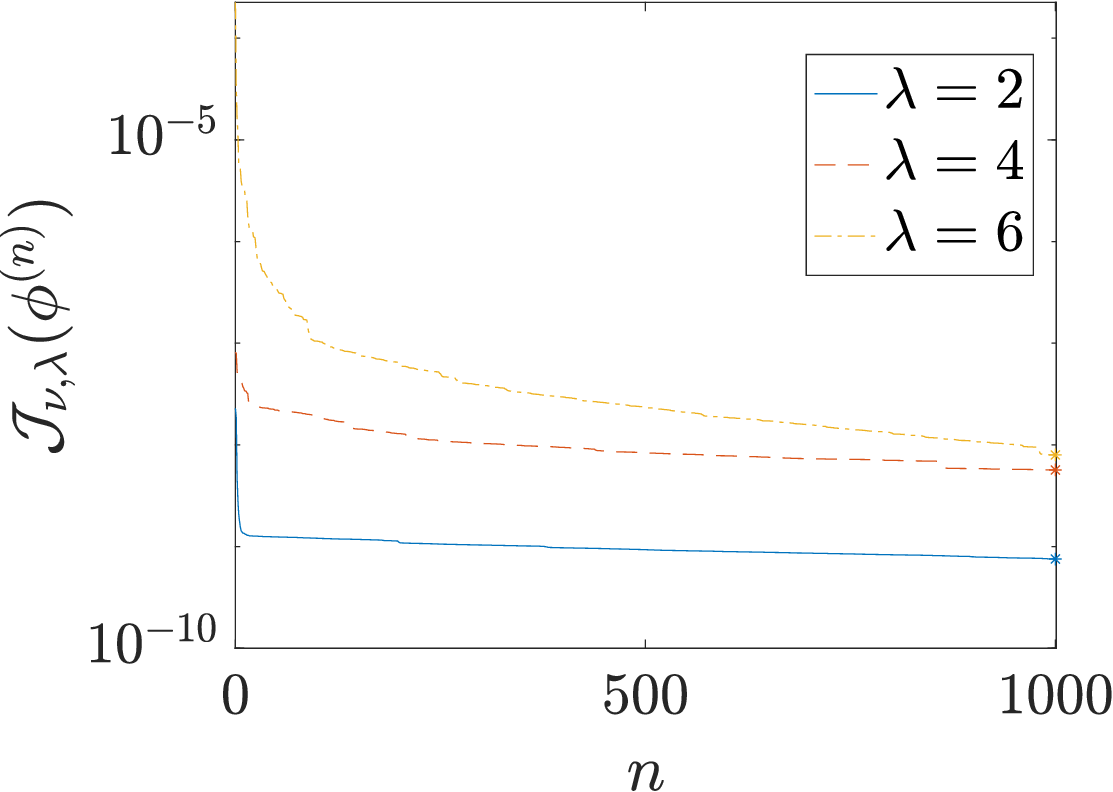}
    \label{fig:sampJ}
  }\qquad
  \subfigure[]
  {
    \includegraphics[scale=0.4]{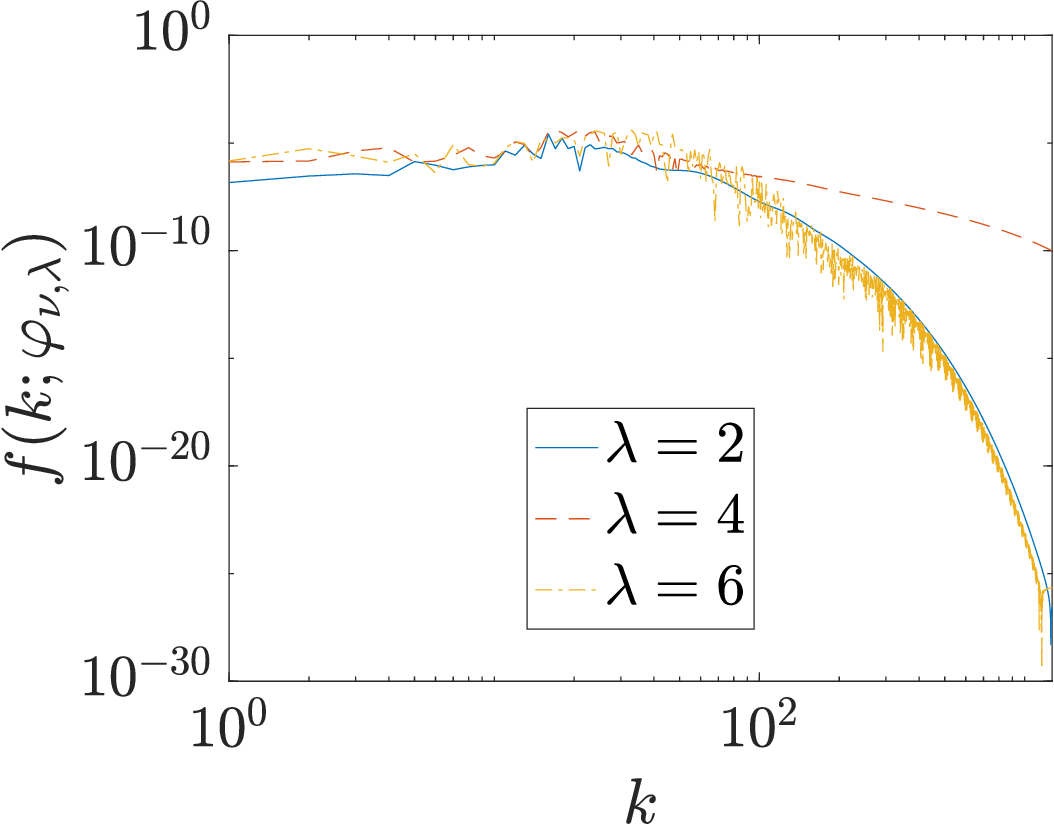}
    \label{fig:sampFourIntG}
  }
  \caption{(a) Decrease of the objective functional \cref{eq:J} with iterations $n$ and (b) the corresponding distributions $f(k;\optphi)$ obtained at the minima when solving \Cref{pb:1} with $\nu = 2.5 \times 10^{-5}$ and (blue solid line) $\lambda = 2$, (red dashed line) $\lambda = 4$, and (yellow dot-dashed line) $\lambda = 6$.
}
\label{fig:Jdecrease}
\end{figure}

\begin{figure}\centering\vspace{-2.2cm}
 \subfigure[]
  {
    \includegraphics[scale=0.4]{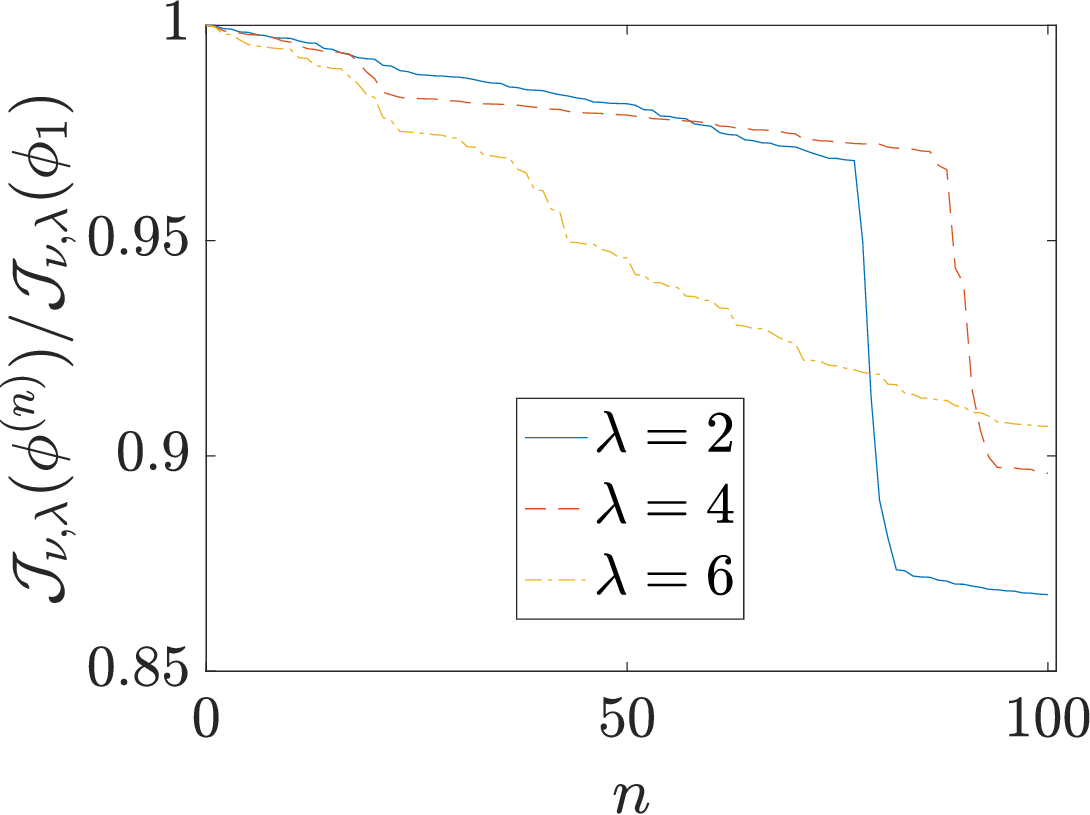}
    \label{fig:Jrefine}
  }\qquad
  \subfigure[]
  {
    \includegraphics[scale=0.4]{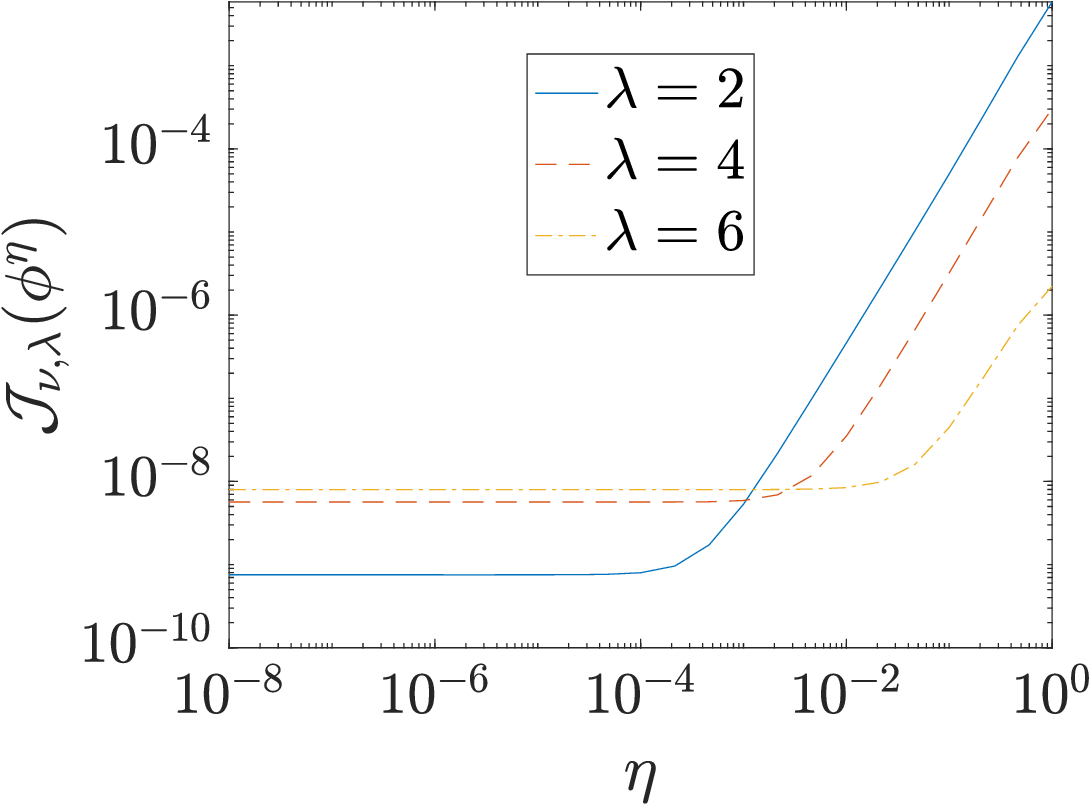}
    \label{fig:Jpert}
  }\qquad
  \subfigure[]
  {
    \includegraphics[scale=0.4]{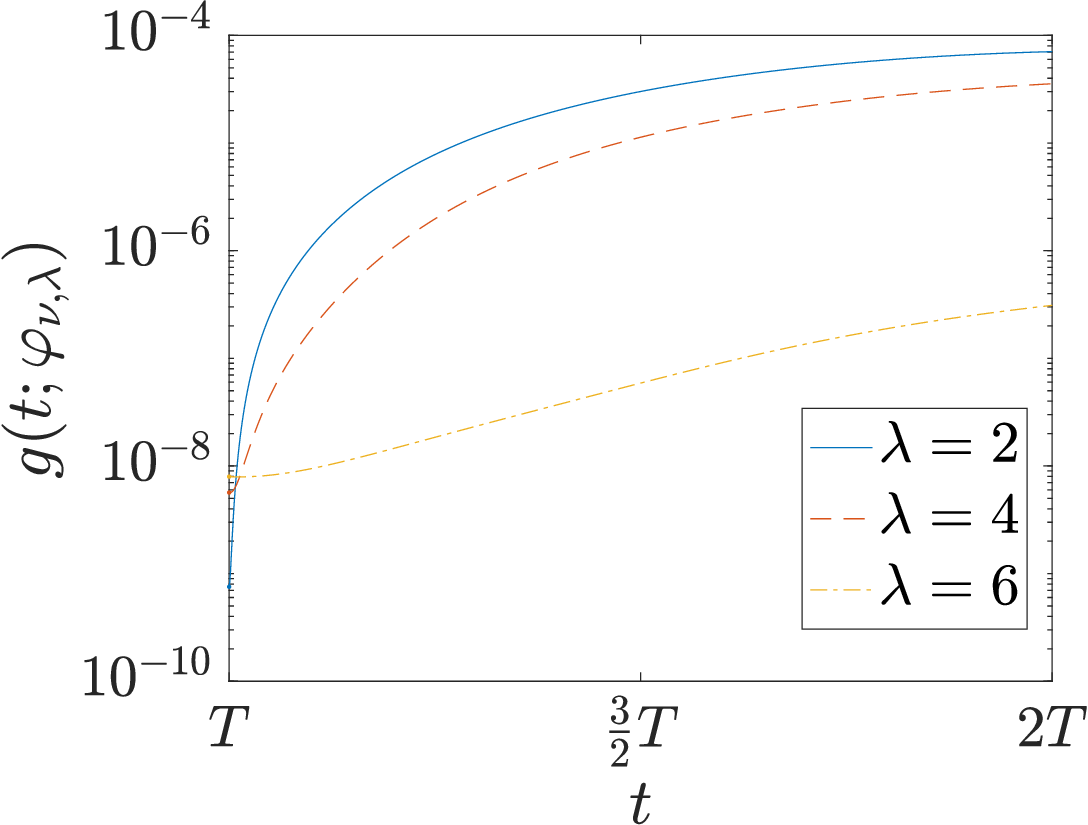}
    \label{fig:JSlide}
  }
      \caption{\revtt{(a) Decrease of the objective functional \cref{eq:J} with iterations $n$ after refinement of the numerical resolution, (b) dependence of the objective functional $\Jl(\phi^{\eta})$ on the noise magnitude $\eta$ in the perturbed initial condition \eqref{eq:phieta}, and \revt{(c) the error function $g(t;\optphi)$ defined over a sliding time window for $t \in [T,2T]$} for $\nu = 2.5 \times 10^{-5}$ and (blue solid line) $\lambda = 2$, (red dashed line) $\lambda = 4$, and (yellow dot-dashed line) $\lambda = 6$.}}
\label{fig:Jrobust}
\end{figure}

\begin{figure}\centering\vspace{-2cm}
  \subfigure[]
  {
    \includegraphics[scale=0.28]{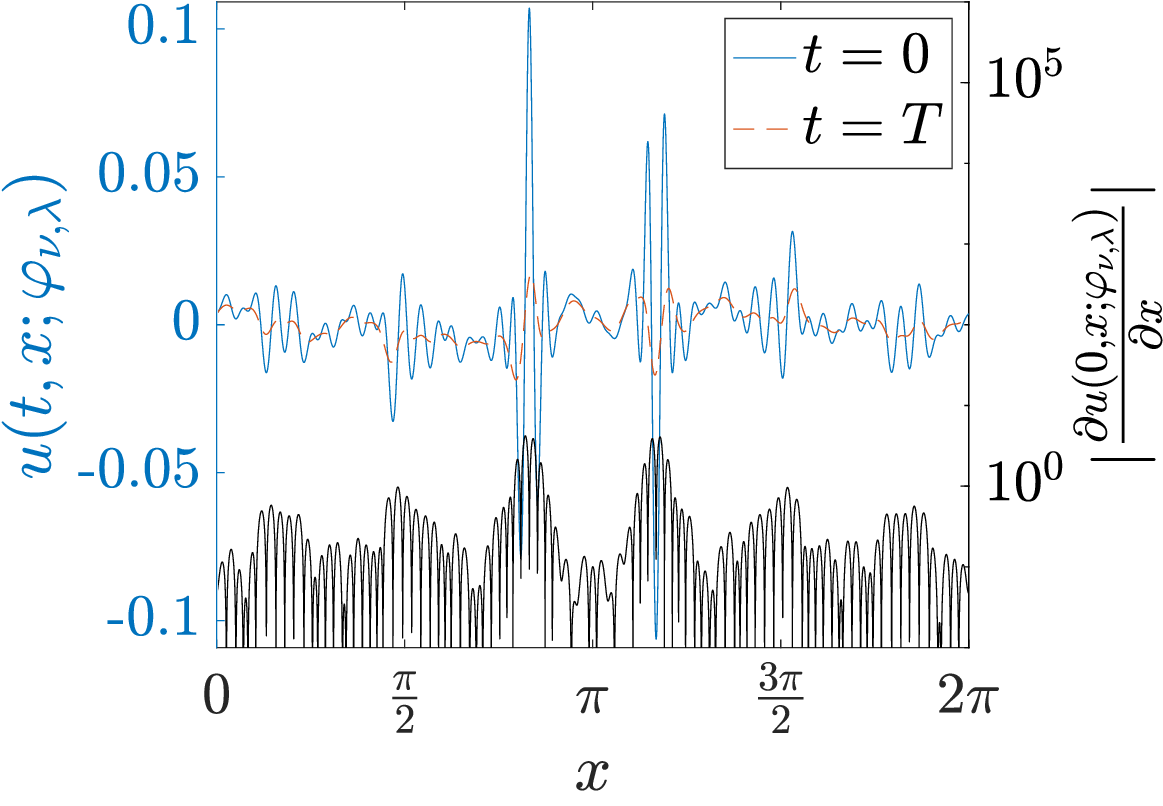}
    \label{fig:ViscPhys}
  }\qquad
  \subfigure[]
  {
    \includegraphics[scale=0.28]{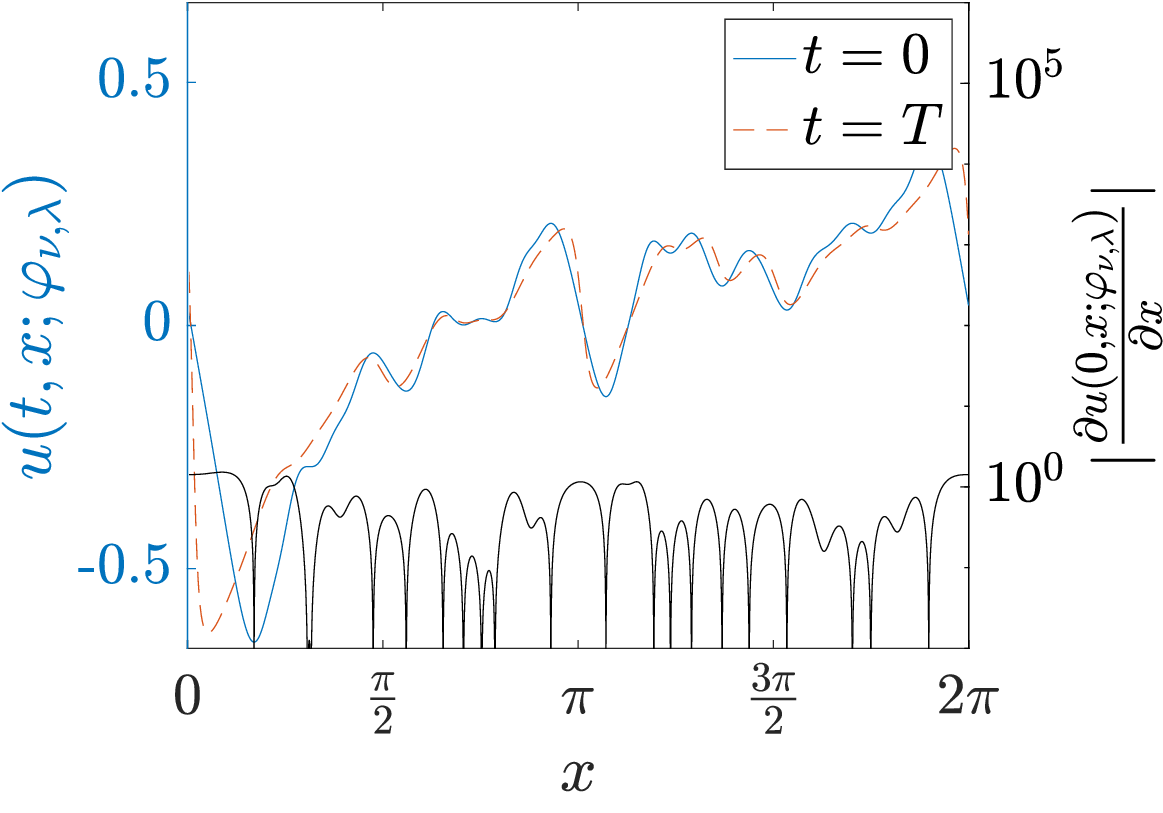}
    \label{fig:InPhys}
  }\qquad
  \subfigure[]
  {
    \includegraphics[scale=0.28]{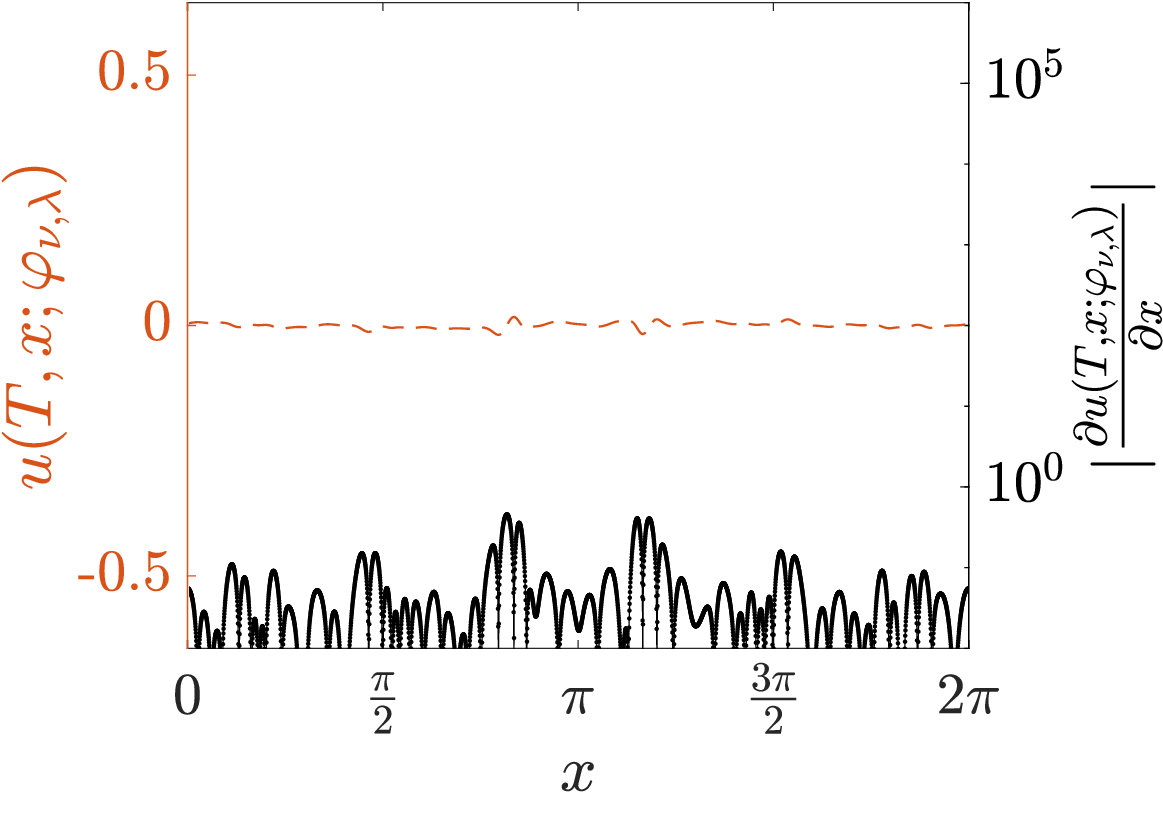}
    \label{fig:Viscdudx}
  }\qquad
  \subfigure[]
  {
    \includegraphics[scale=0.28]{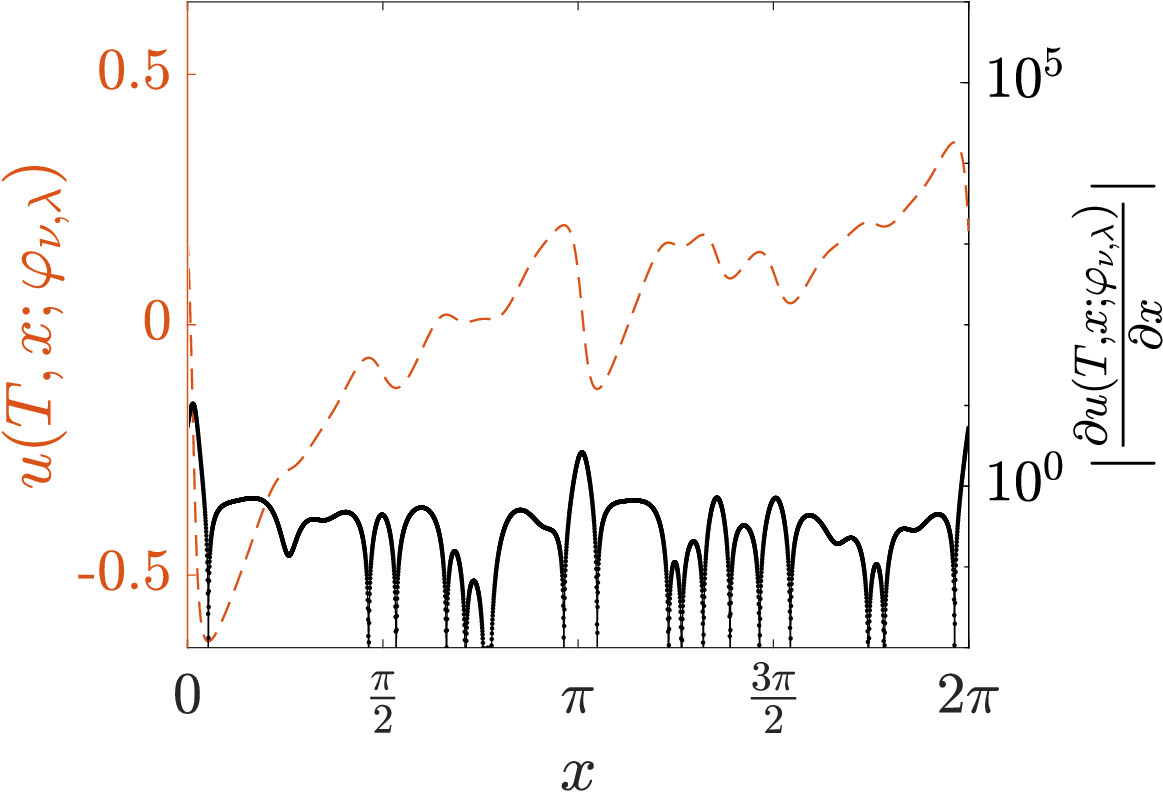}
    \label{fig:Indudx}
  }\qquad
  \subfigure[]
  {
    \includegraphics[scale=0.28]{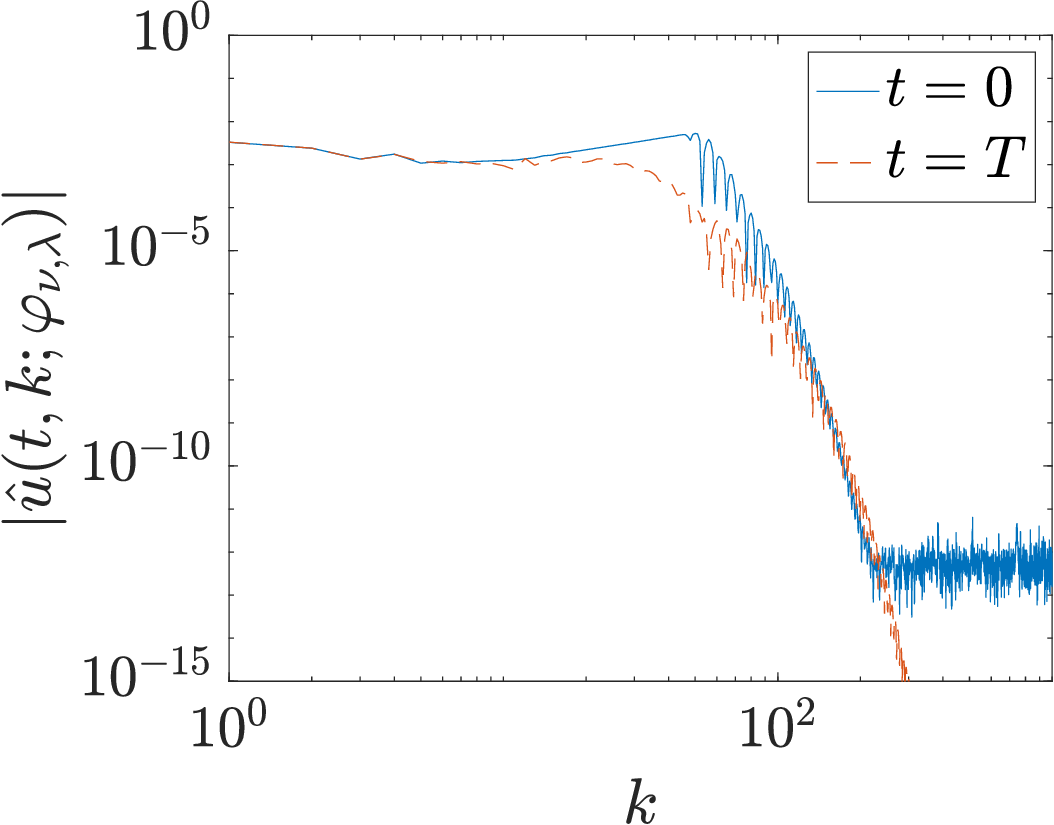}
    \label{fig:ViscFour0T}
  }\qquad
  \subfigure[]
  {
    \includegraphics[scale=0.28]{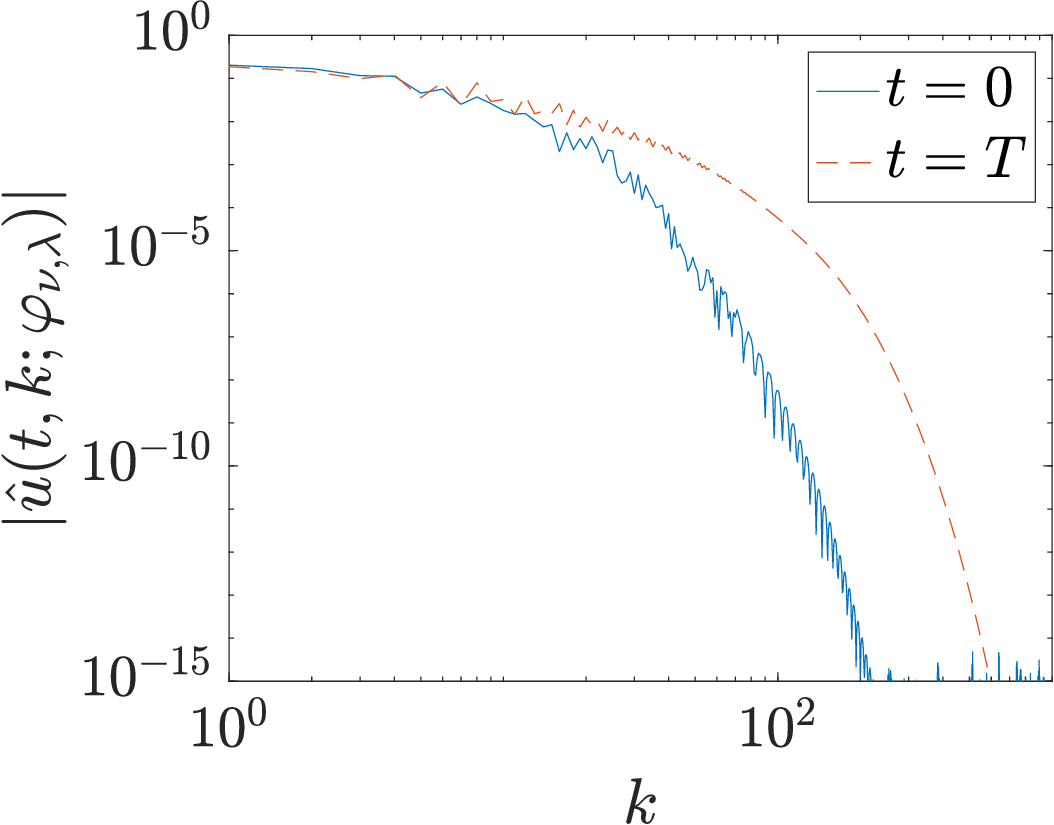}
    \label{fig:InFour0T}
  }\qquad
  \subfigure[]
  {
    \includegraphics[scale=0.28]{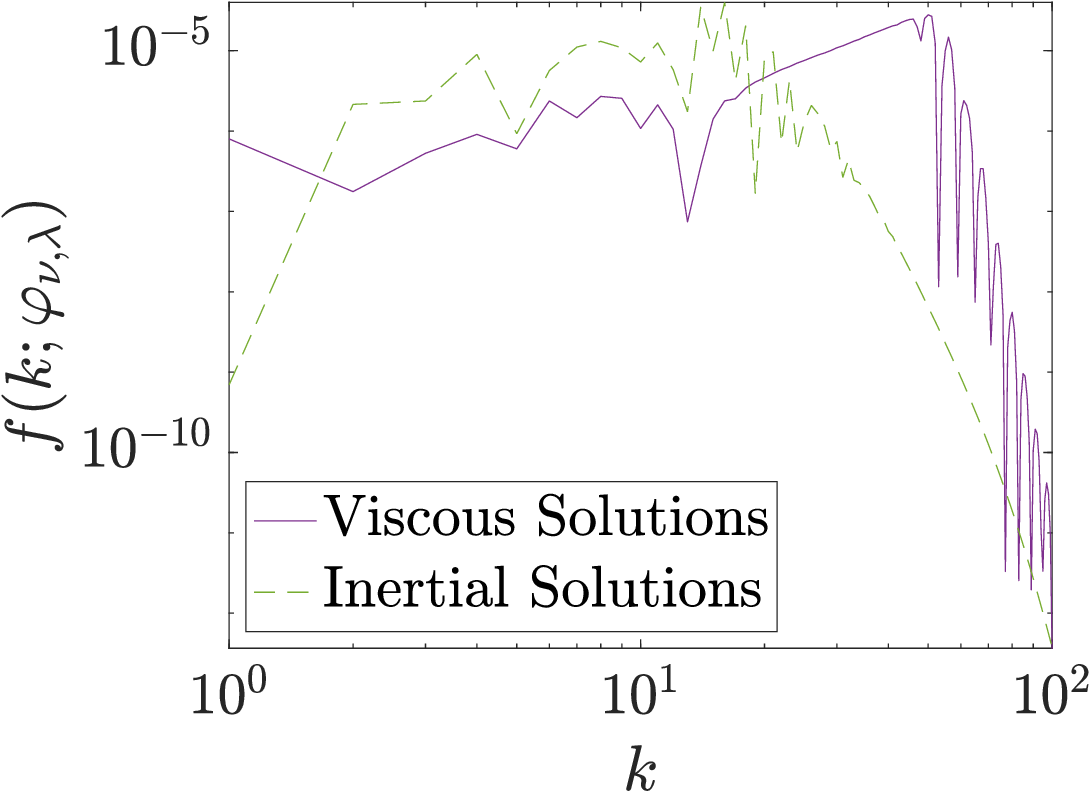}
    \label{fig:compFourIntF}
  }\qquad
  \subfigure[]
  {
    \includegraphics[scale=0.28]{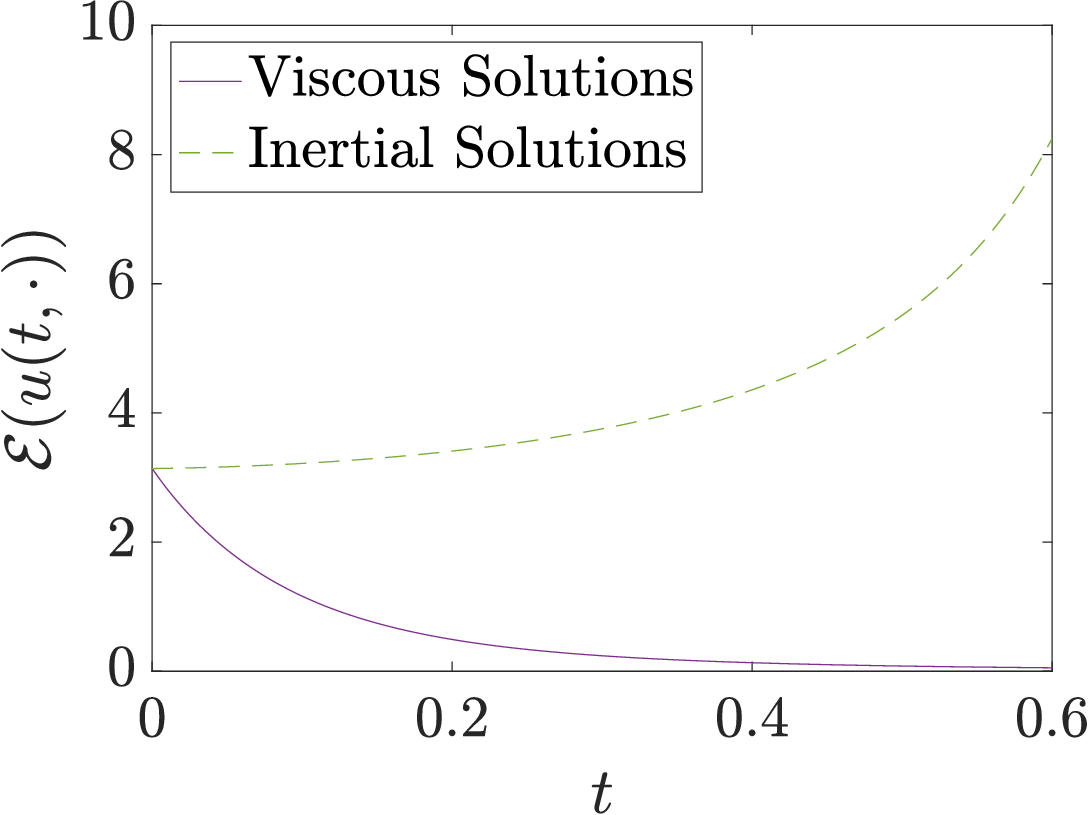}
    \label{fig:compEnst}
  }
  \caption{Comparison of the properties of (left column) the viscous solutions and (right column) the inertial solutions obtained by solving \Cref{pb:1} with $\nu = 2.5 \times 10^{-3}$ and $\lambda=2$: (a,b) optimal initial conditions $\optphi(x)$ (blue, left axis), the magnitude of their gradients $\left|d\optphi(x)/d x \right|$ (black, right axis) and the corresponding solutions $u(T,x;\optphi)$ at the final time (red, left axis); (c,d) the solutions at the final time $u(T,x;\optphi)$ (red, left axis) and their gradients $\left|\partial u(T,x;\optphi) / \partial x \right|$ (black, right axis); (e,f) the Fourier spectra $\widehat{\optphi}(k)$ (blue) and $\widehat{u}(T,k;\optphi)$ (red) of the optimal initial conditions and of the corresponding final states; (g, h) the functions $f(k;\optphi)$ vs.~$k$ and $\E(u(t, \cdot;\optphi))$ vs.~$t$, cf.~\eqref{eq:J} and \eqref{eq:E}, respectively, at the minimum for the viscous solution (purple) and the inertial solution (green).}
\label{fig:ViscInComp}
\end{figure}

\setlength\tabcolsep{0.5pt}
\begin{table}
	\centering\vspace{-2cm}
	\vspace*{-0.5cm} \hspace*{-0.5cm}\begin{tabular}{ |c|c|c|c|c|} 
		\hline
		\rowcolor{Gray}
		  & 
\centering $\optphi(x)$
& \centering $u(T, x; \optphi)$  
& \centering $\widehat{\varphi}_{\nu, \lambda}(k)$
& \begin{minipage}{0.15 \textwidth}
 \centering $\widehat{u}(T, k; \optphi)$
\end{minipage} ~\\ 
		\hline \\ [-1.5em]
		\rotatebox{90}{~$\mathbf{\nu = 2.5 \times 10^{-3}}$} 
&{\includegraphics[scale=0.22]
{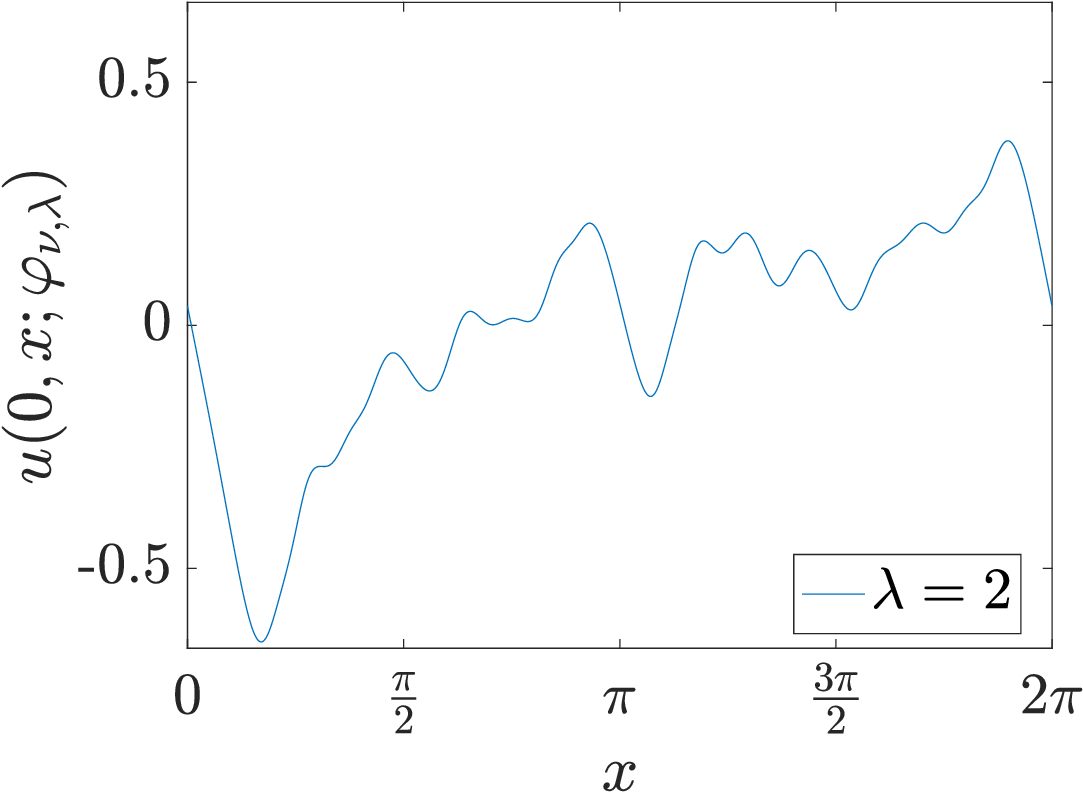}} 
&{\includegraphics[scale=0.22]{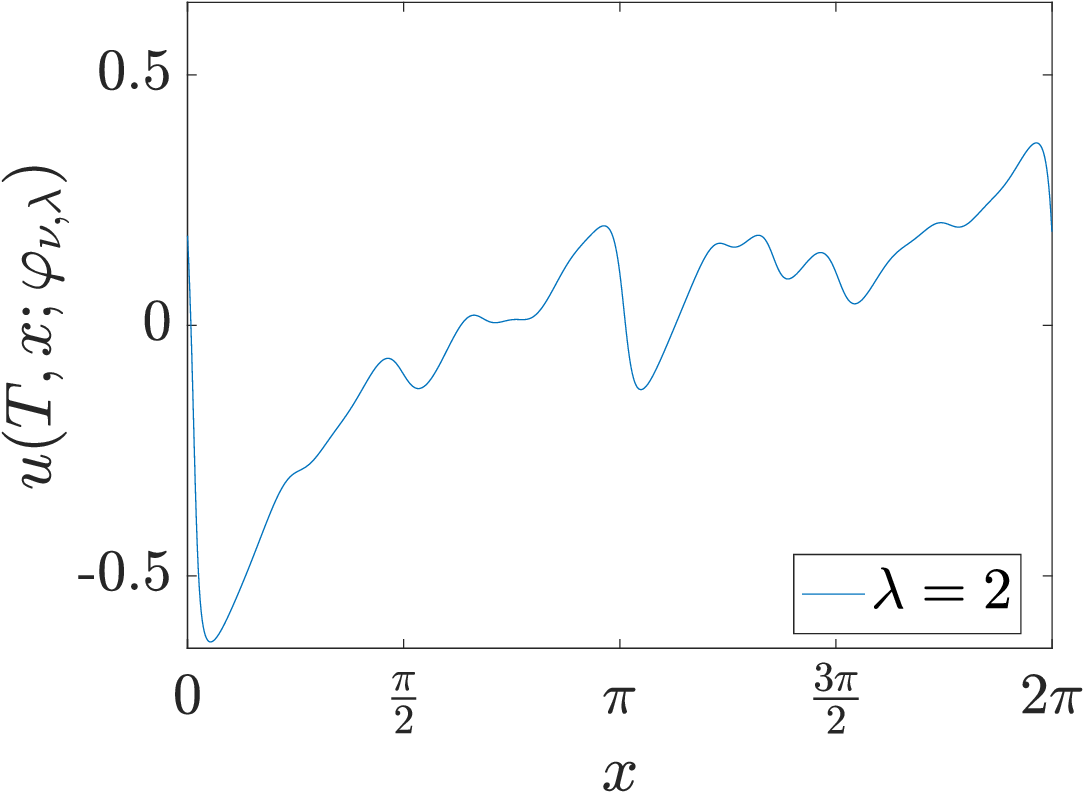}} 
&{\includegraphics[scale=0.22]{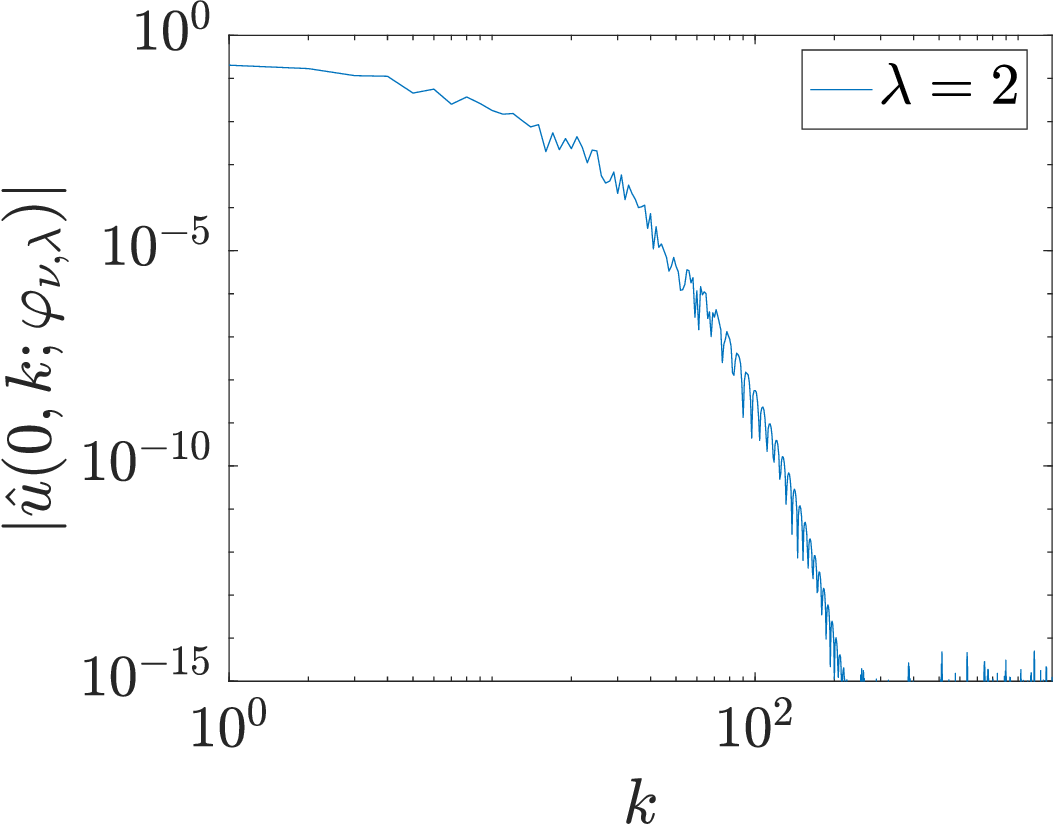}} 
&{\includegraphics[scale=0.22]{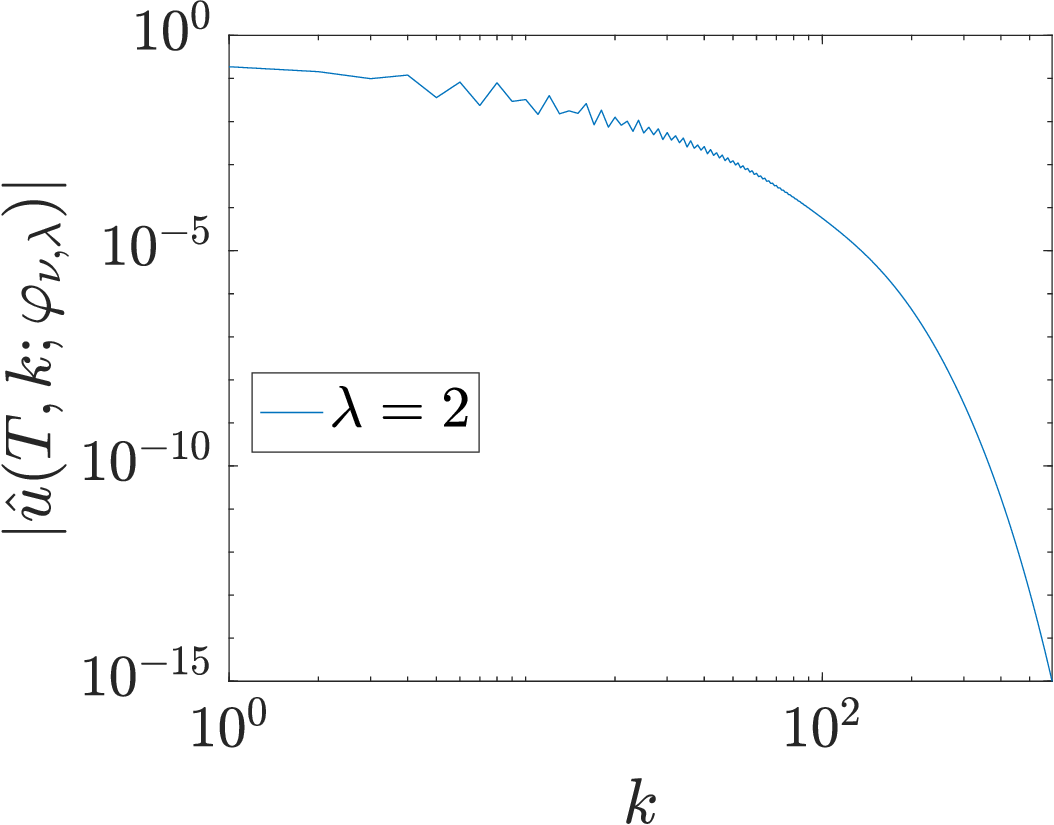}} 
		\\		
		\hline \\ [-1.5em]
		\rotatebox{90}{~$\mathbf{\nu = 2.5 \times 10^{-4}}$} 
& {\includegraphics[scale=0.22]{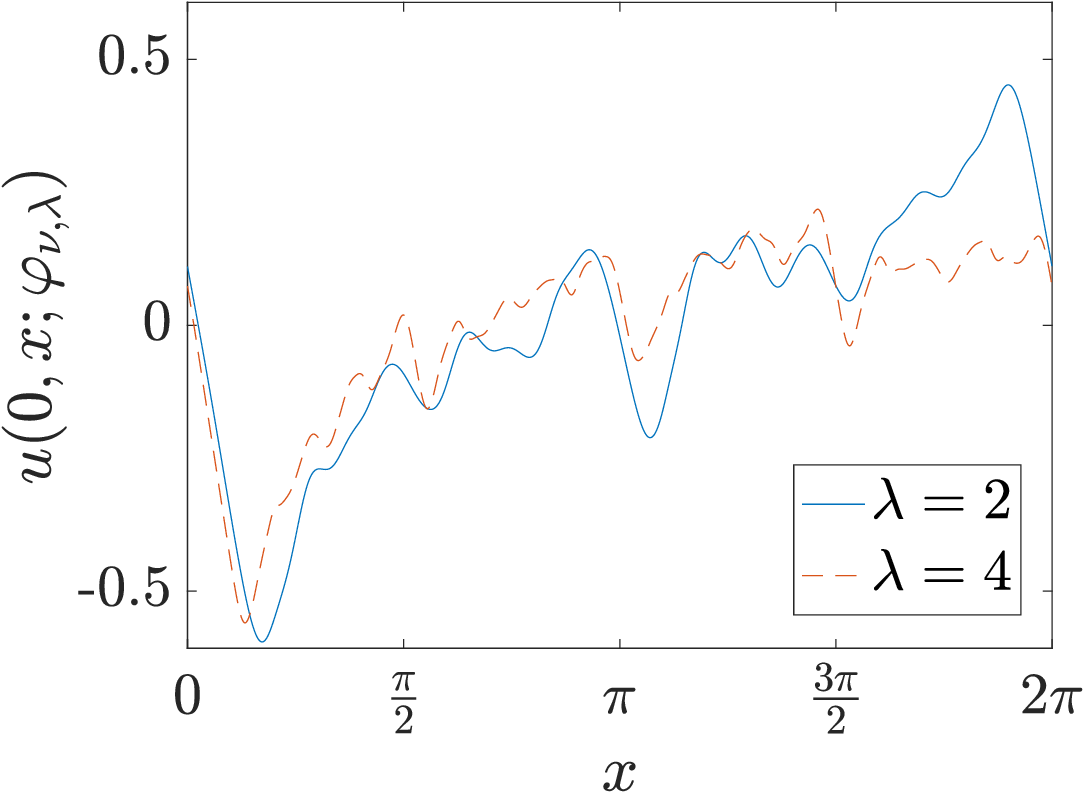}}
&{\includegraphics[scale=0.22]{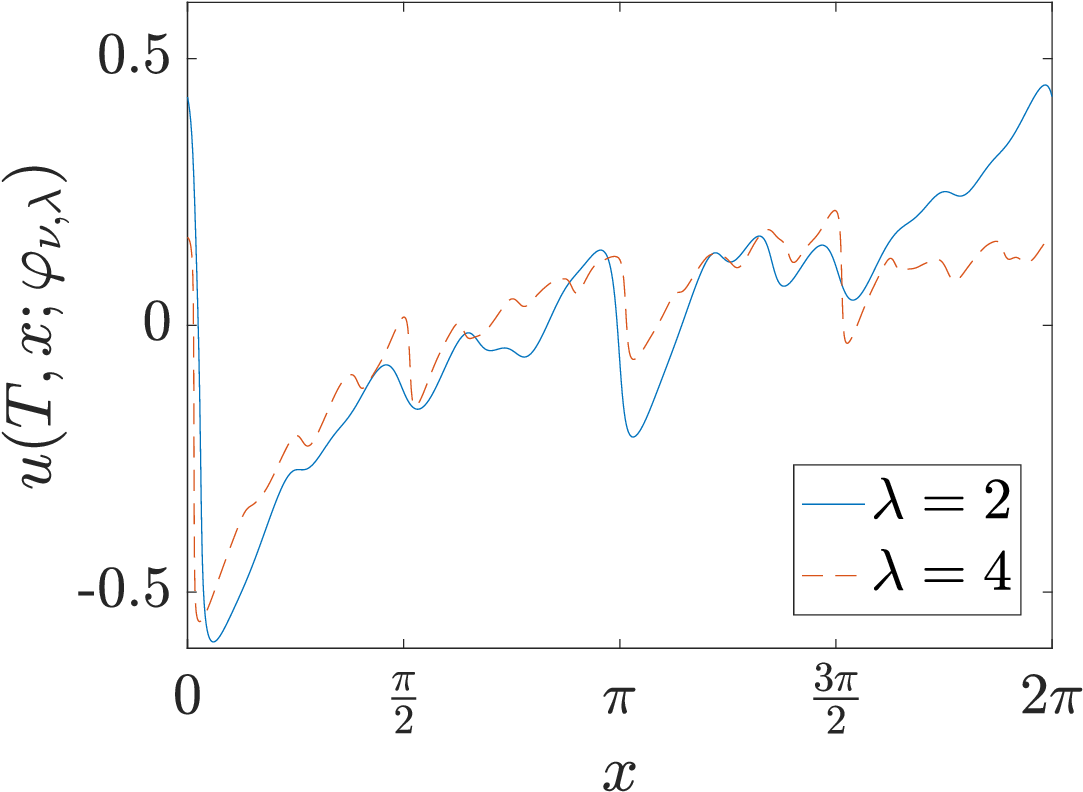}}
&{\includegraphics[scale=0.22]{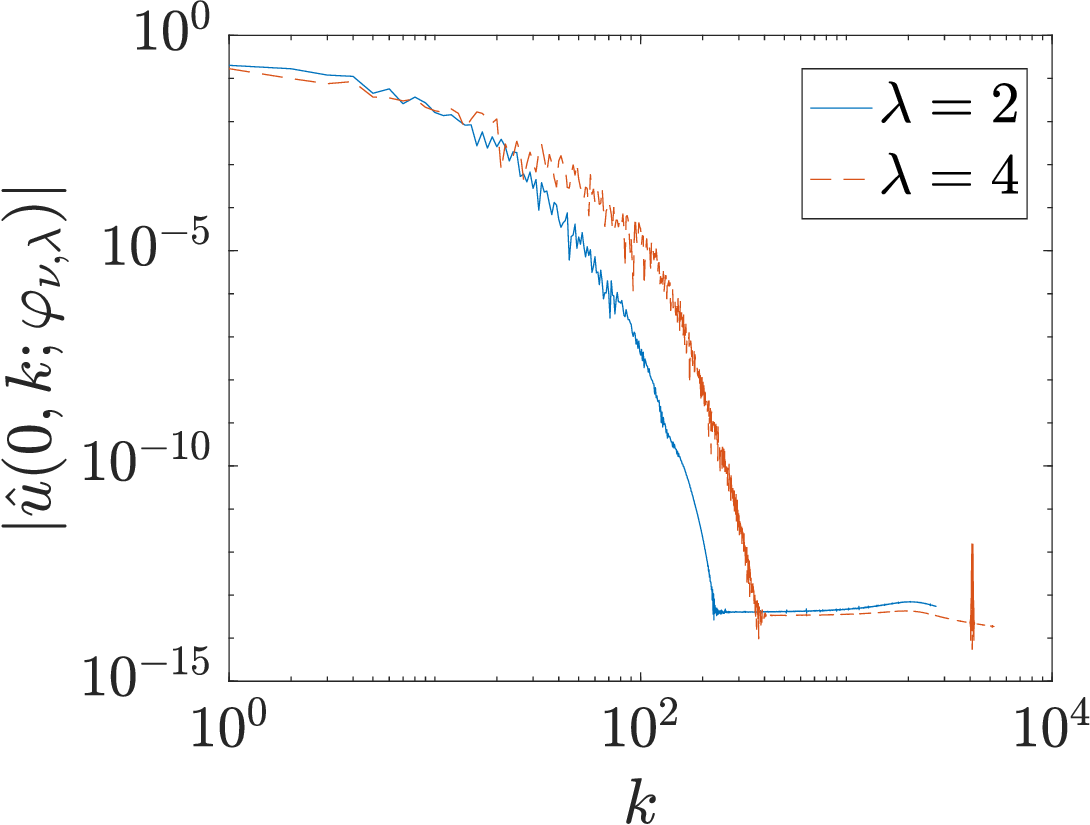}}
&{\includegraphics[scale=0.22]{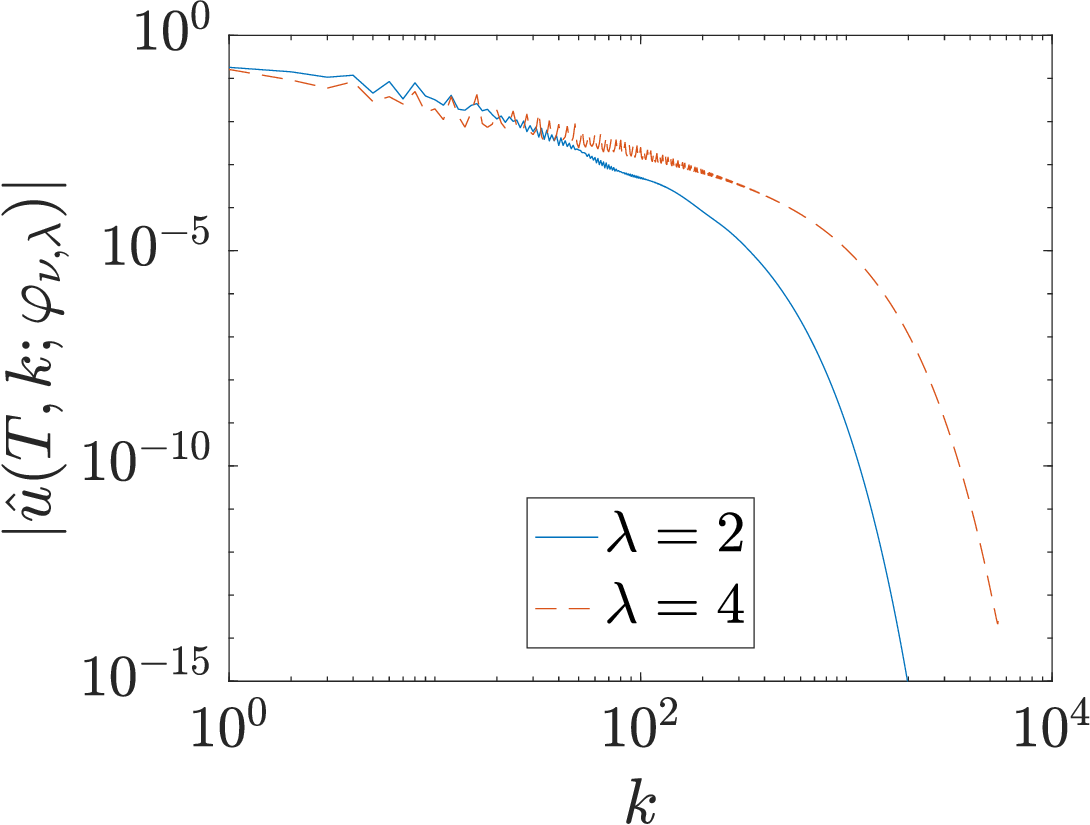}} 
	\\		
		\hline \\ [-1.5em]
		\rotatebox{90}{~$\mathbf{\nu = 2.5 \times 10^{-4}}$} 
& {\includegraphics[scale=0.22]{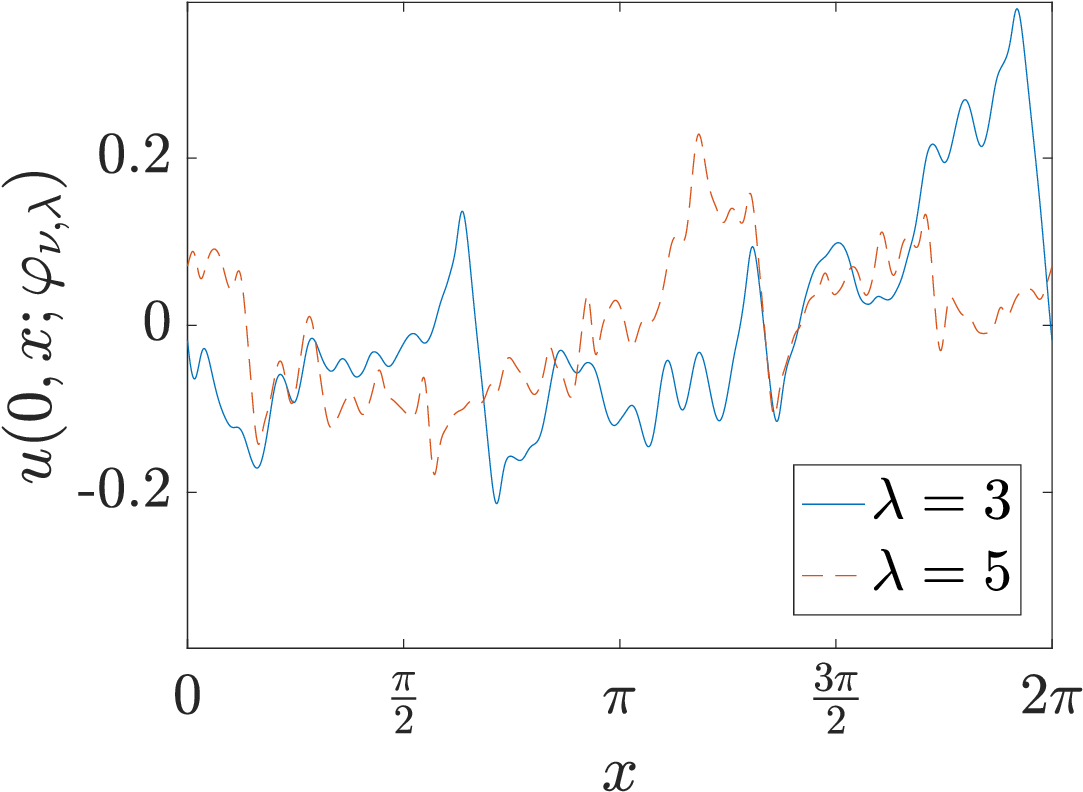}}
&{\includegraphics[scale=0.22]{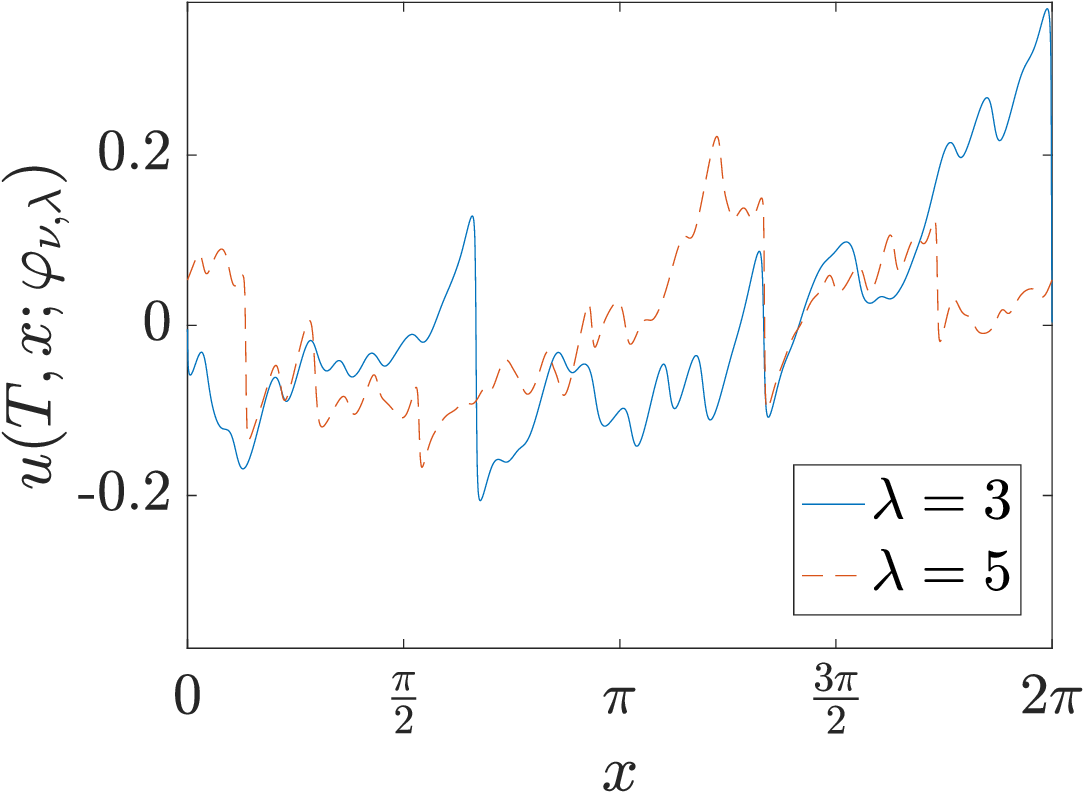}}
&{\includegraphics[scale=0.22]{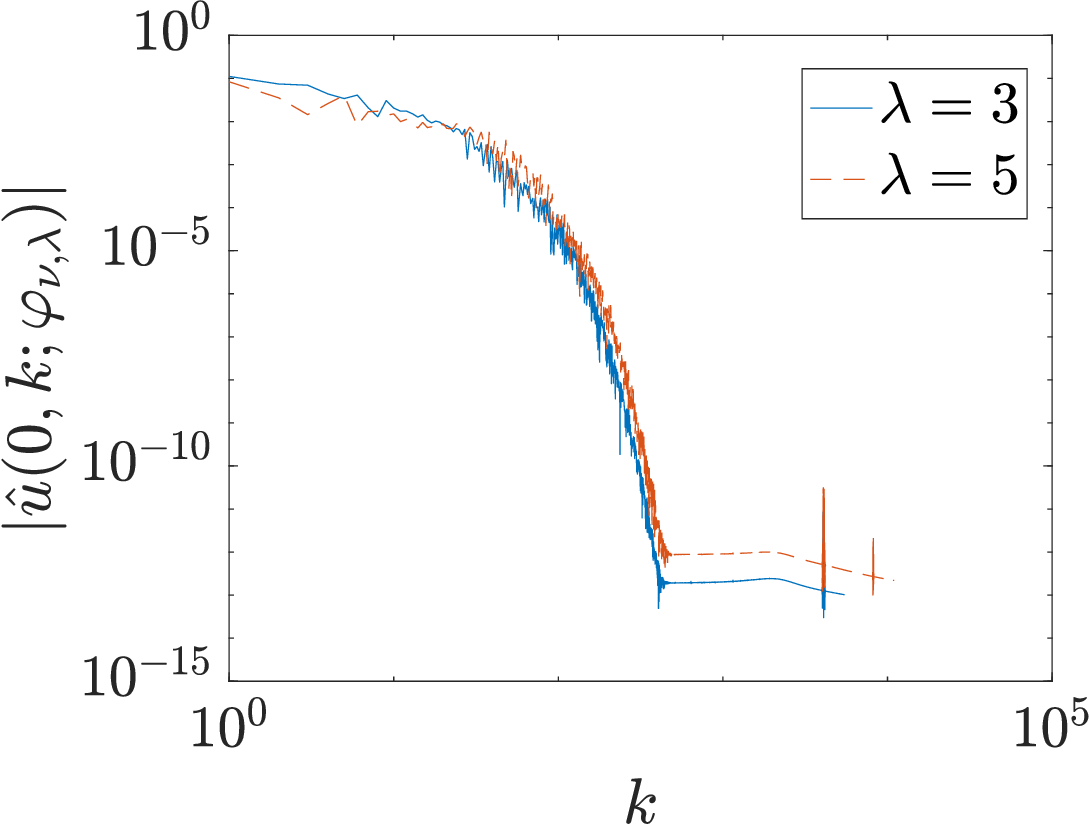}}
&{\includegraphics[scale=0.22]{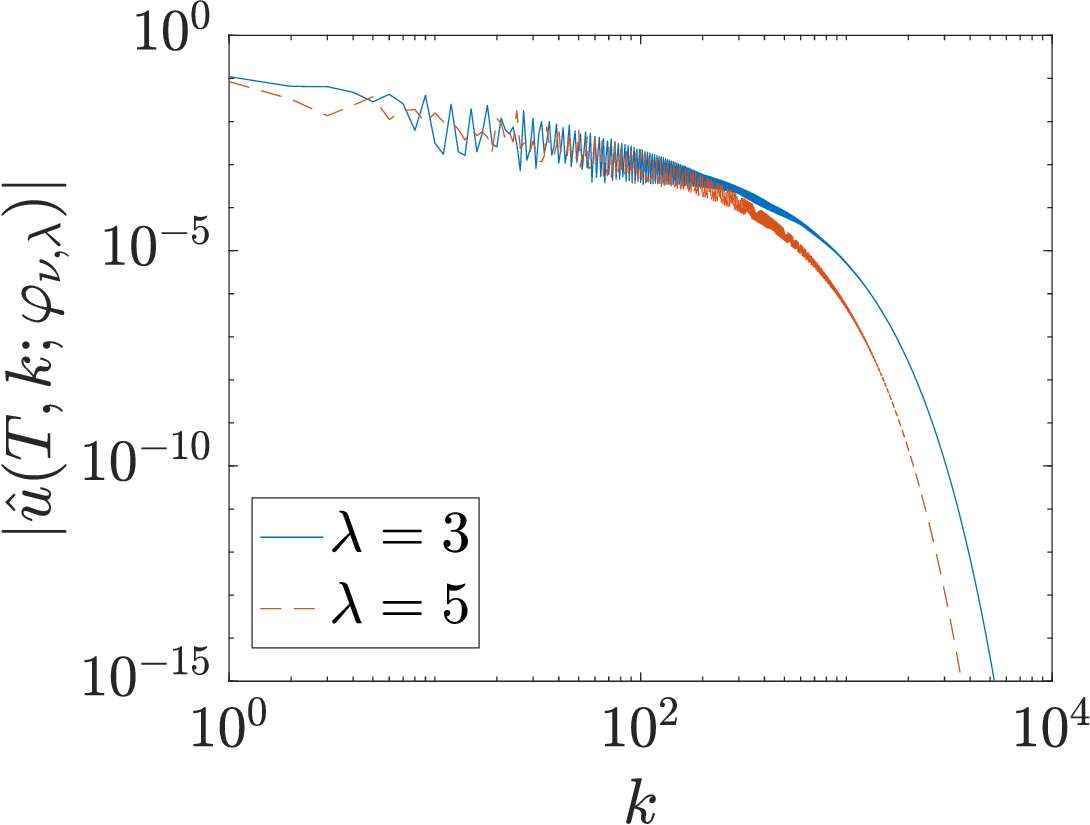}} 
		\\		
		\hline \\ [-1.5em]
		\rotatebox{90}{~$\mathbf{\nu = 2.5 \times 10^{-5}}$} 
&{\includegraphics[scale=0.22]{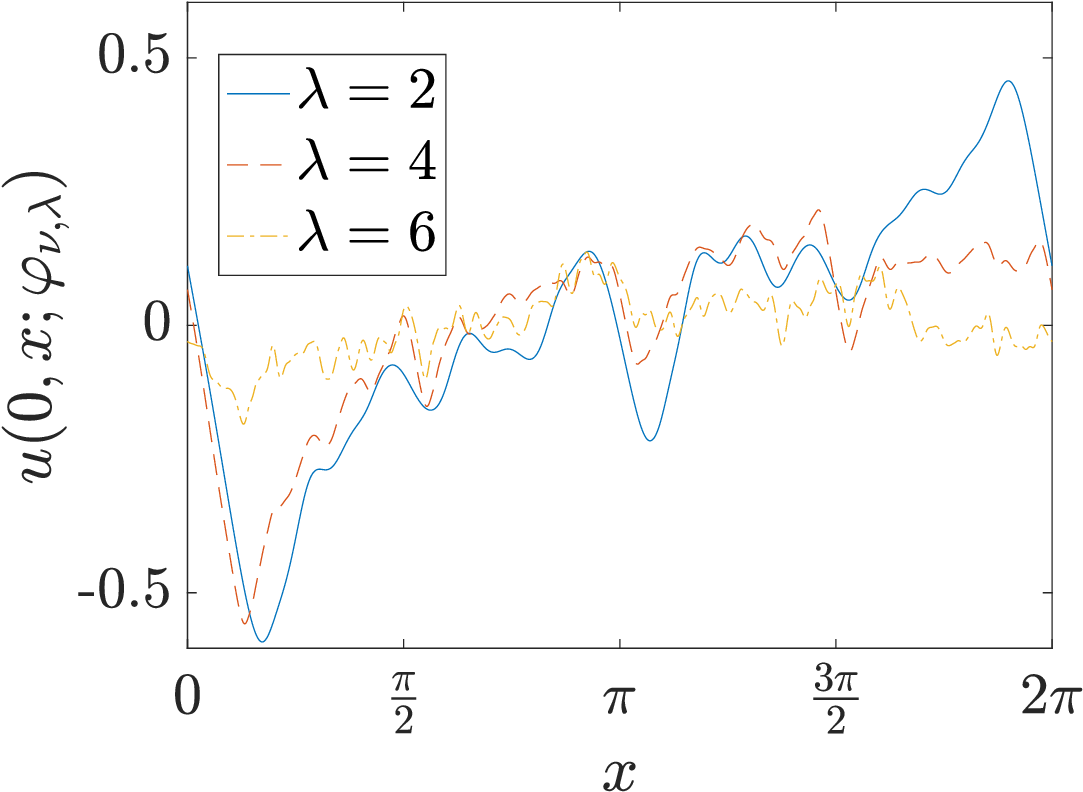}} 
&{\includegraphics[scale=0.22]{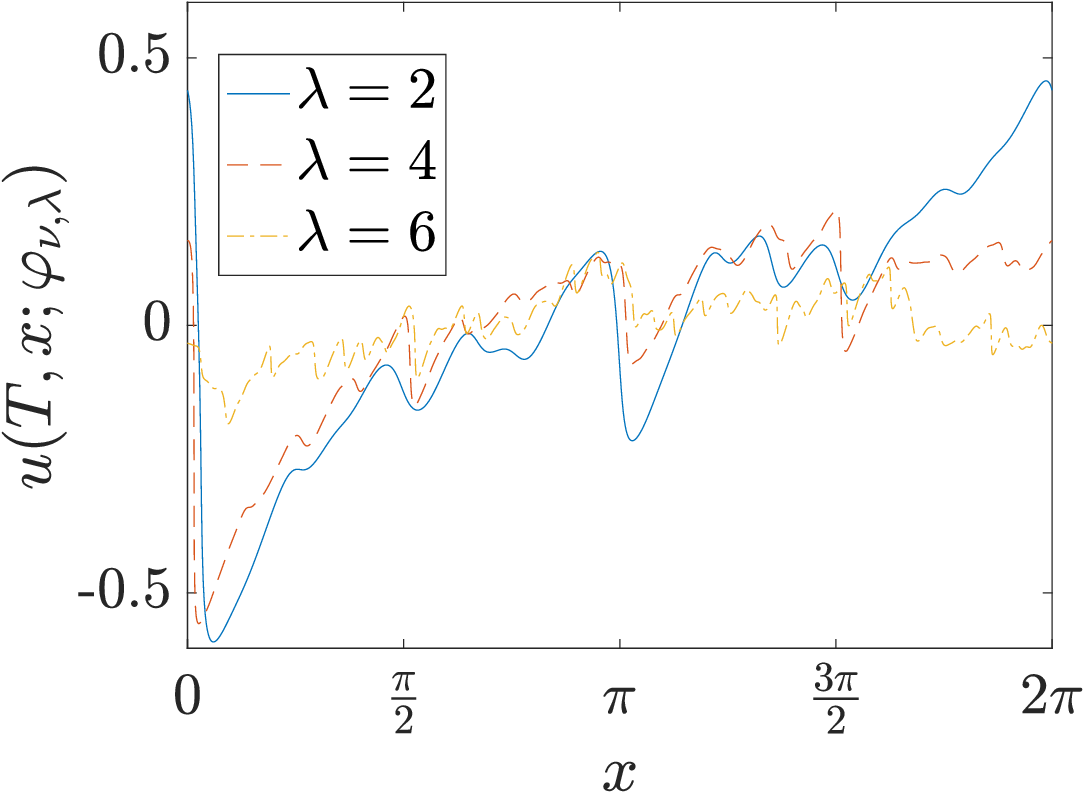}} 
&{\includegraphics[scale=0.22]{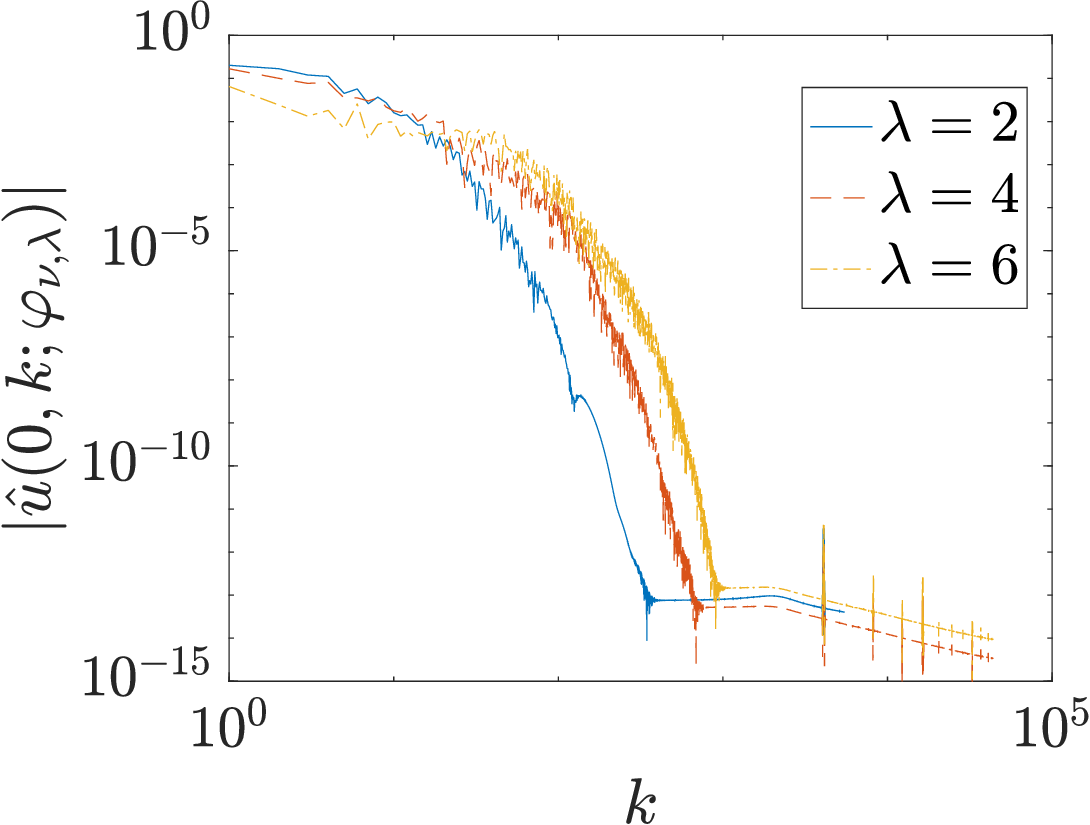}} 
&{\includegraphics[scale=0.22]{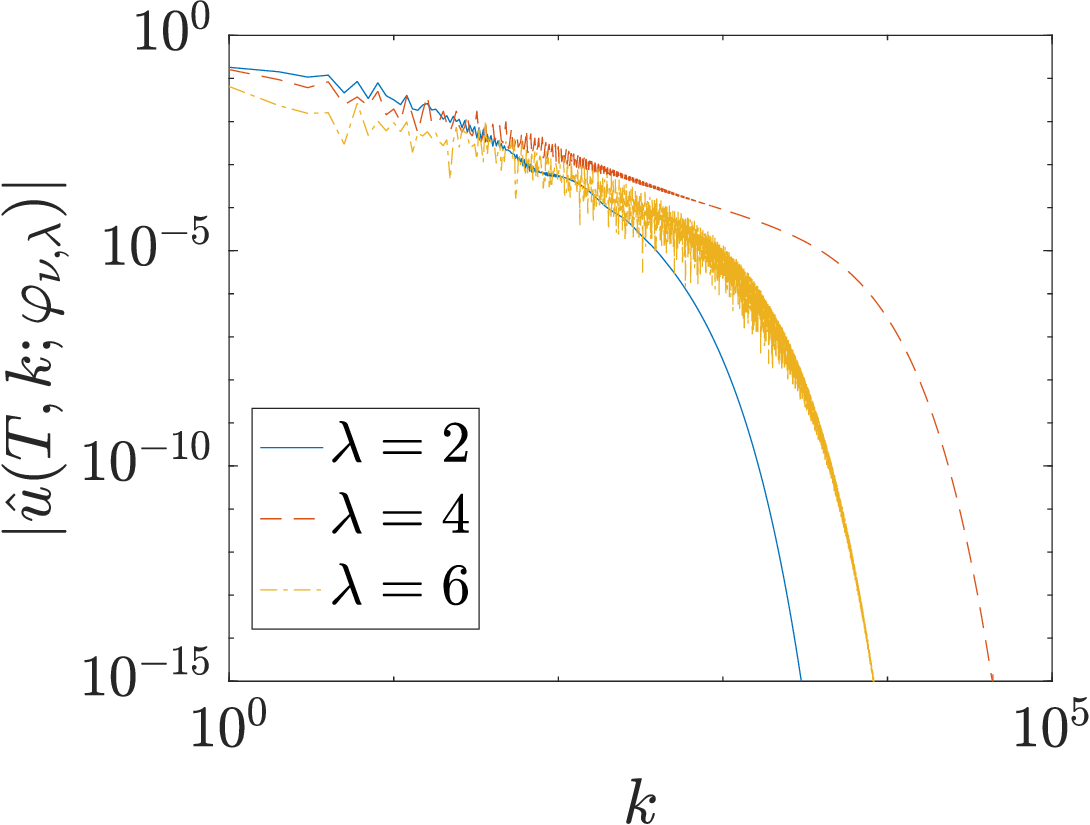}} 
		\\		
		\hline \\ [-1.5em]
		\rotatebox{90}{~$\mathbf{\nu = 2.5 \times 10^{-5}}$} 
&{\includegraphics[scale=0.22]{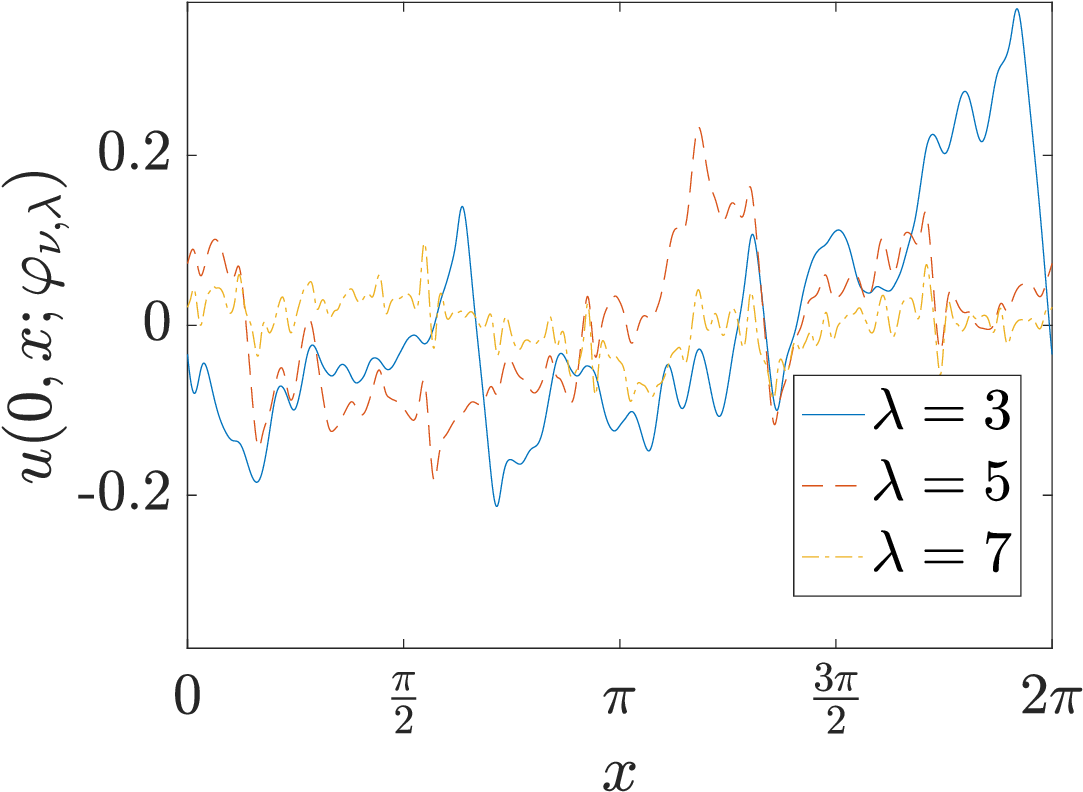}} 
&{\includegraphics[scale=0.22]{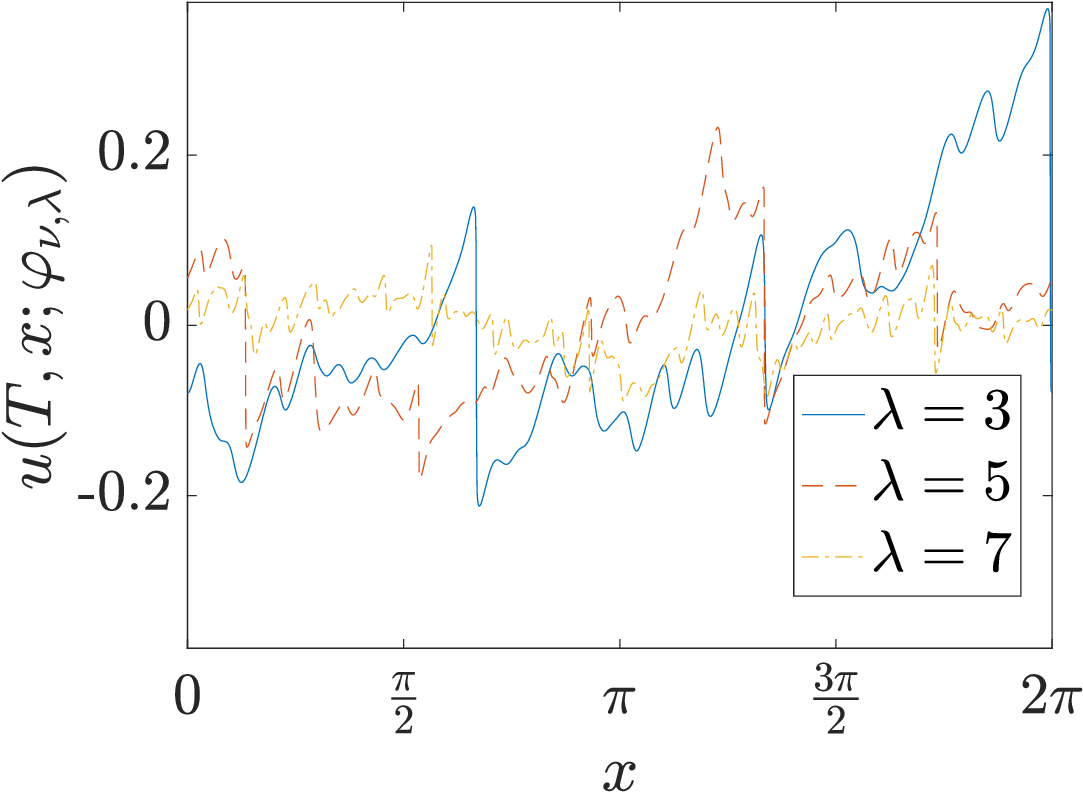}} 
&{\includegraphics[scale=0.22]{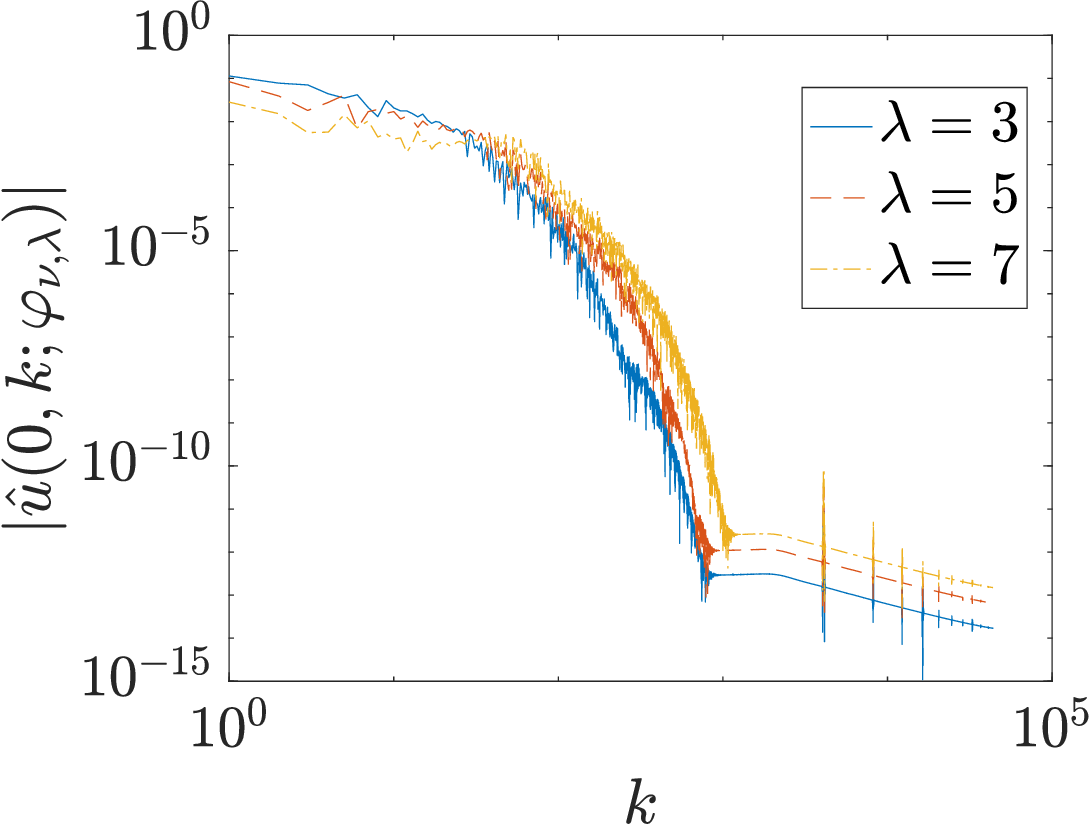}} 
&{\includegraphics[scale=0.22]{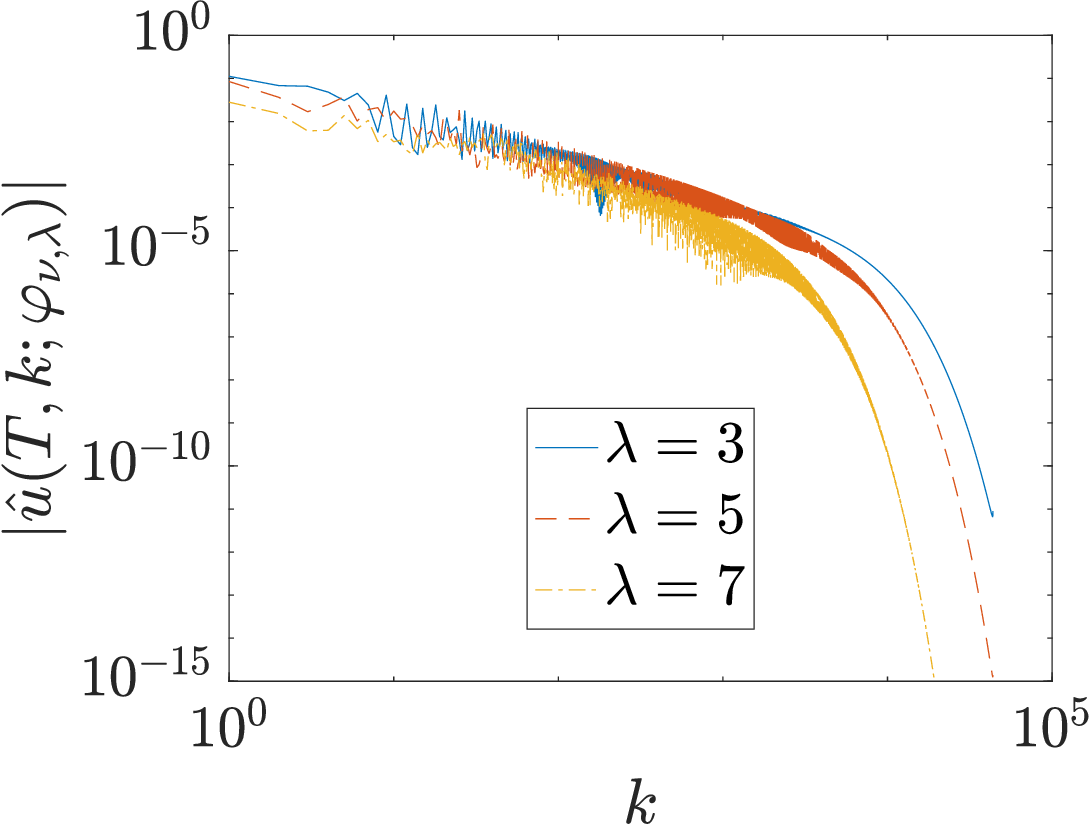}} 
		\\		
		\hline
	\end{tabular}
\caption{Summary information about the inertial solutions found by solving \Cref{pb:1} with the values of $\nu$ indicated on the left and the values of $\lambda$ given in the individual legends. For a better comparison the solutions obtained with even and odd $\lambda$ are grouped together. The first and the second columns represent the optimal initial conditions $\optphi$ and the final states $u(T,x;\optphi)$, with the corresponding spectra given in the third and fourth columns.}
\label{tbl:LambComp}
\end{table} 

\begin{table}
	\centering
	\vspace*{-0.5cm} \hspace*{-0.5cm}\begin{tabular}{ |c|c|c|c|c|} 
		\hline
		\rowcolor{Gray}
		  & 
\centering $\optphi(x)$
& \centering $u(T, x; \optphi)$  
& \centering $\widehat{\varphi}_{\nu, \lambda}(k)$
& \begin{minipage}{0.15 \textwidth}
 \centering $\widehat{u}(T, k; \optphi)$
\end{minipage} \\ 
		\hline \\ [-1.5em]
		\rotatebox{90}{\centering \qquad $\mathbf{\lambda=2}$} 
&{\includegraphics[scale=0.22]
{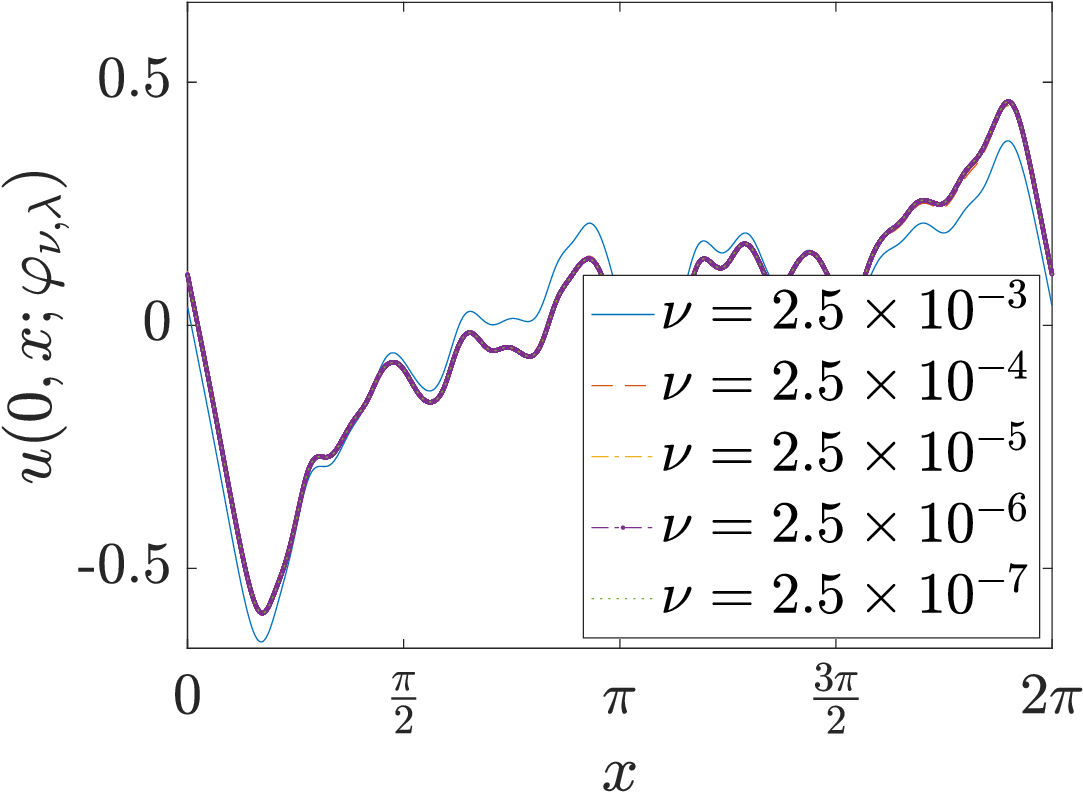}} 
&{\includegraphics[scale=0.22]{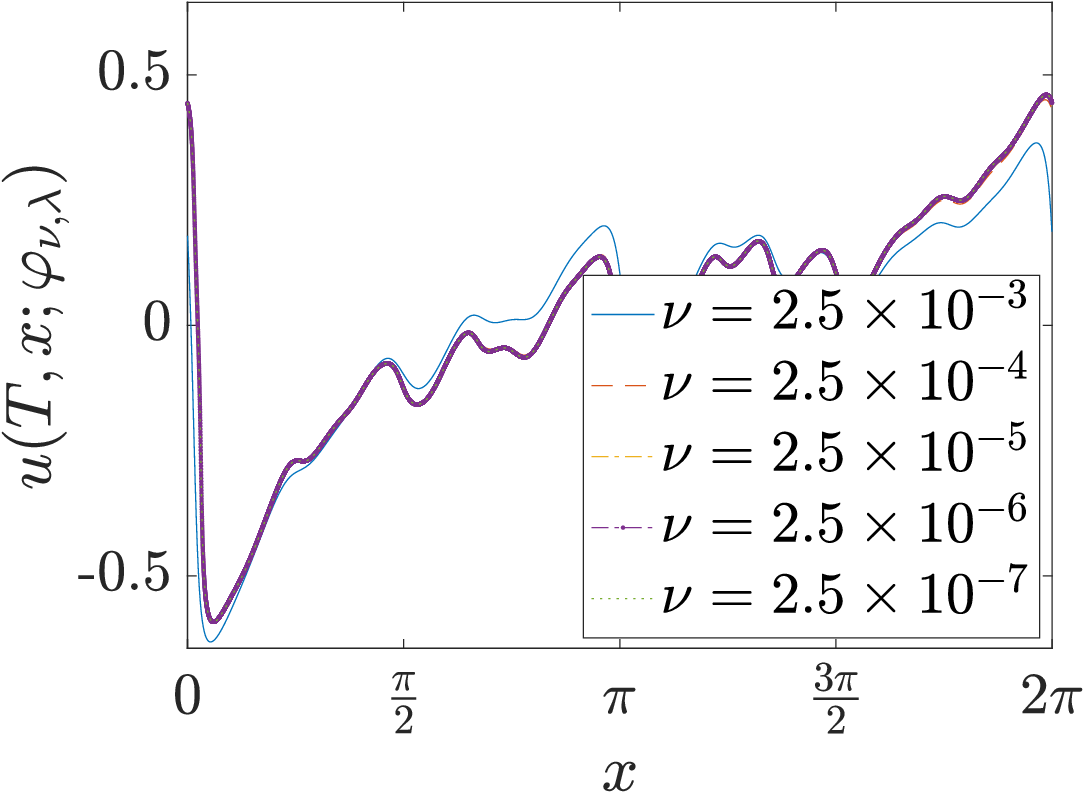}} 
&{\includegraphics[scale=0.22]{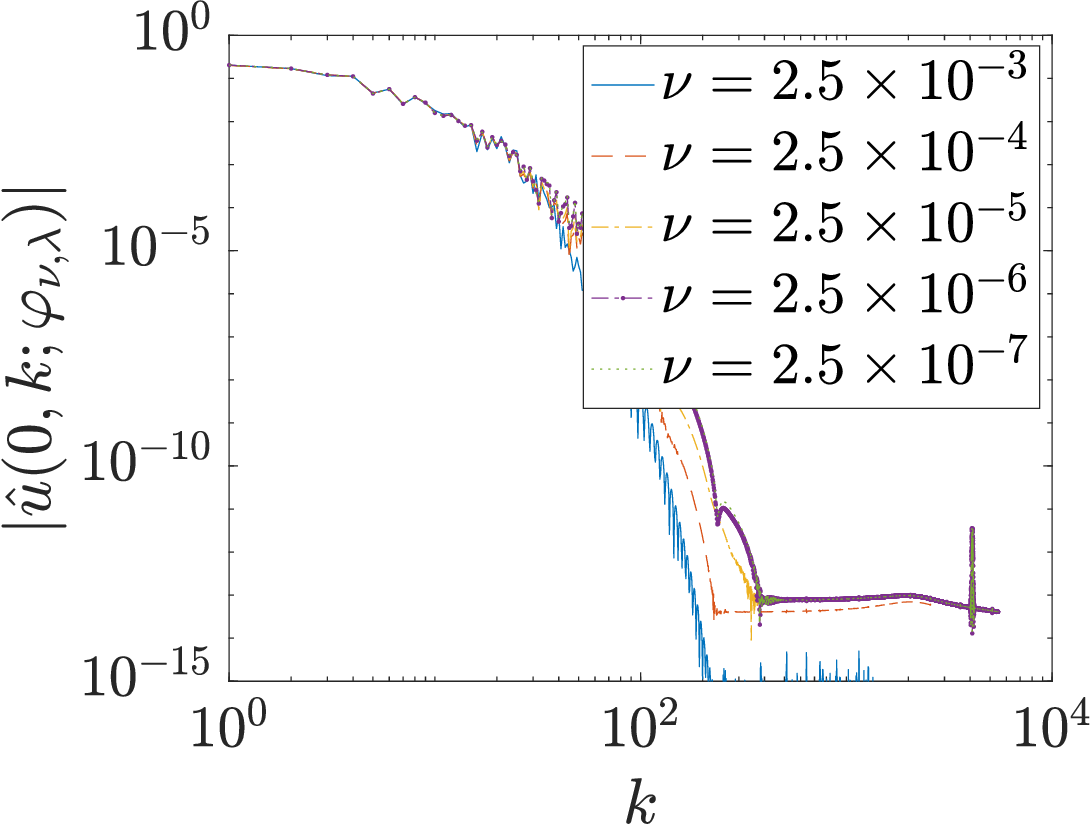}} 
&{\includegraphics[scale=0.22]{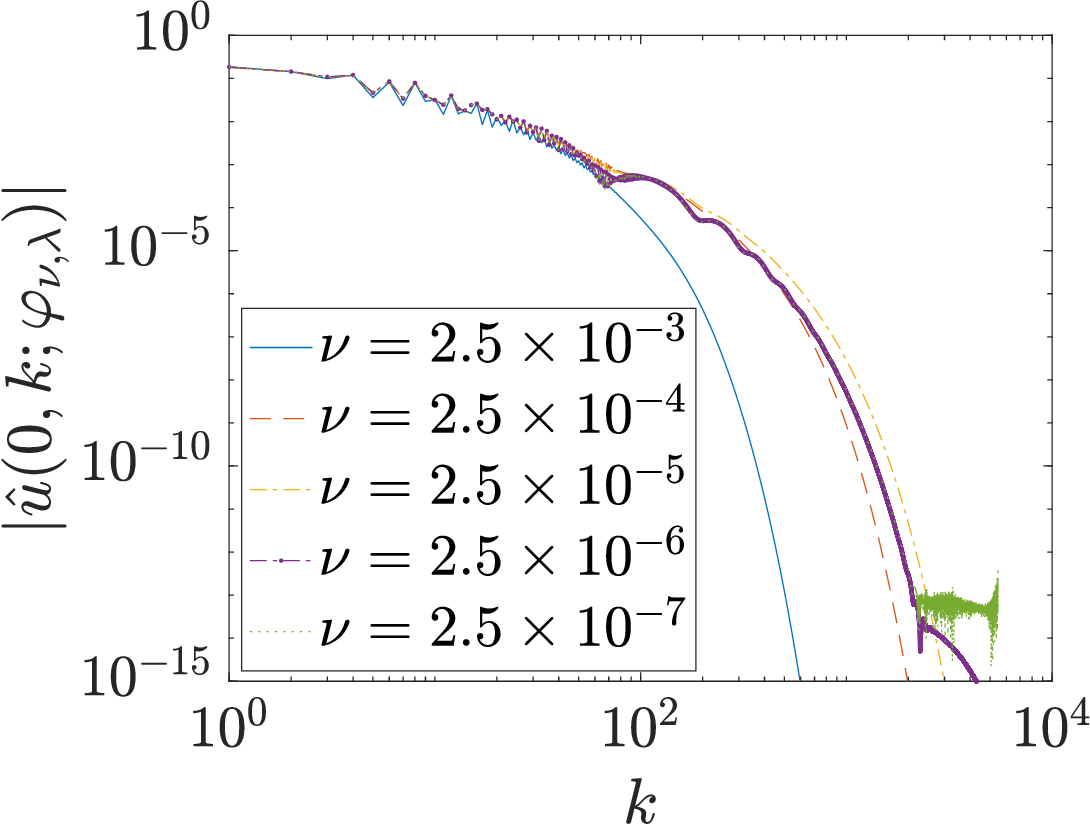}} 
		\\		
		\hline \\ [-1.5em]
		\rotatebox{90}{\centering \qquad $\mathbf{\lambda=5}$} 
& {\includegraphics[scale=0.22]{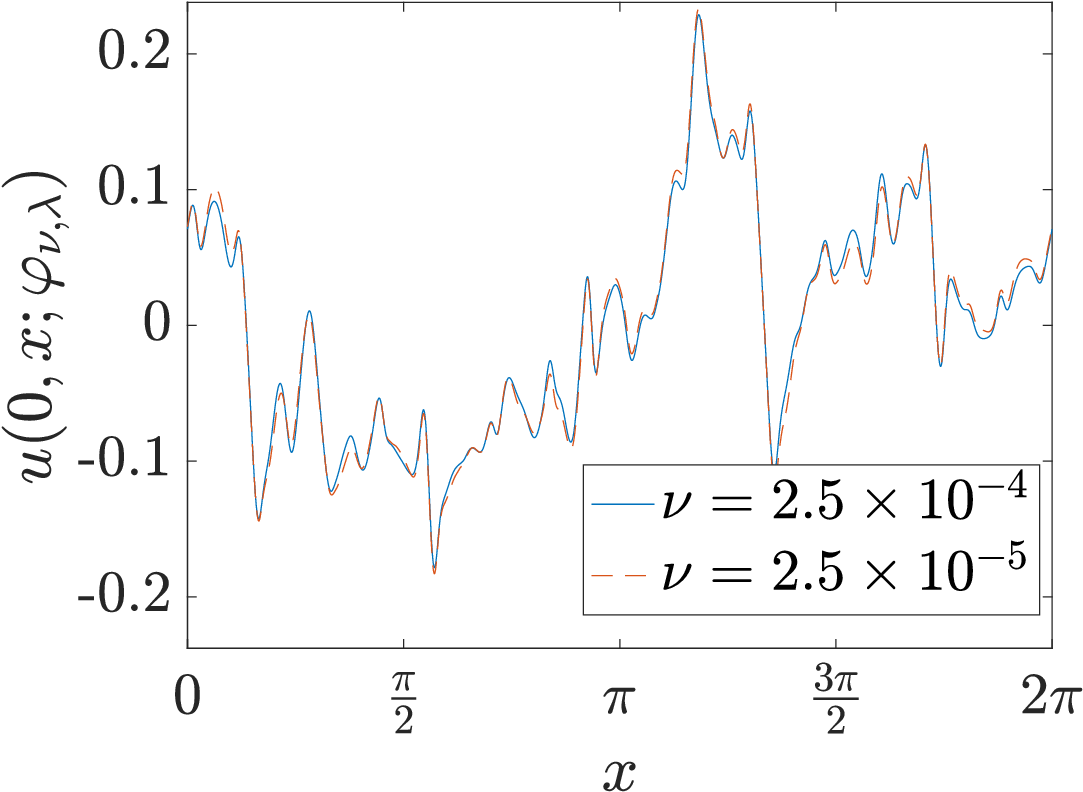}}
&{\includegraphics[scale=0.22]{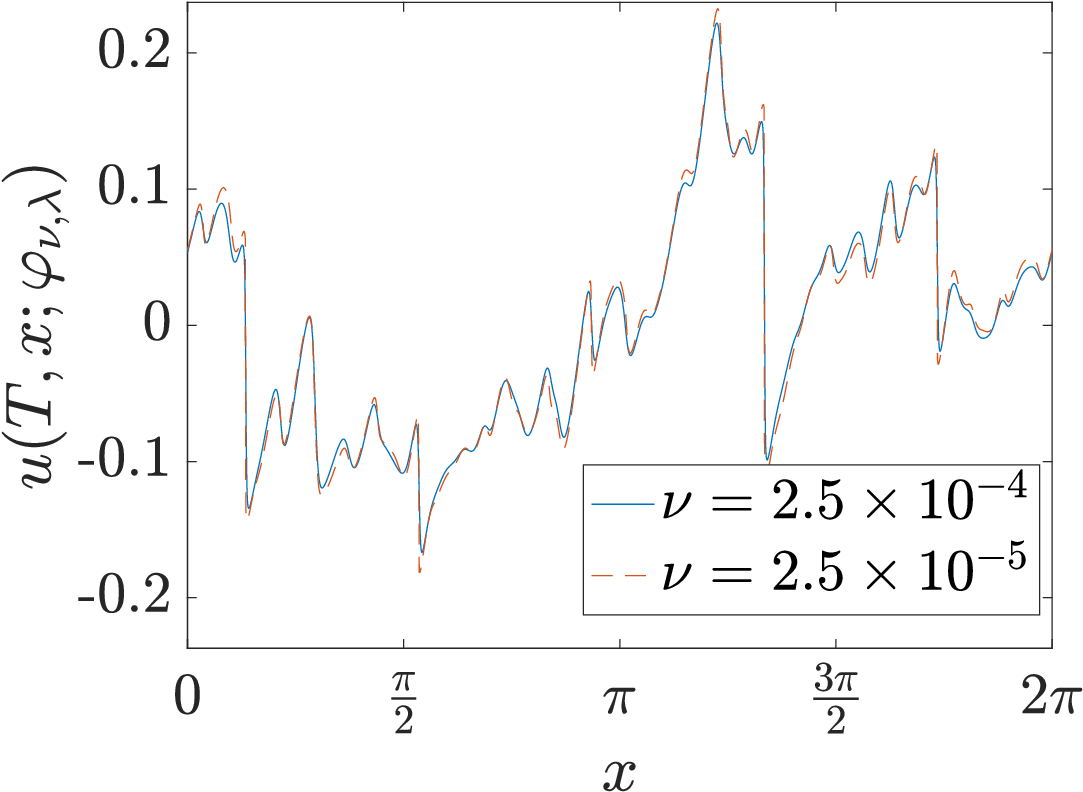}}
&{\includegraphics[scale=0.22]{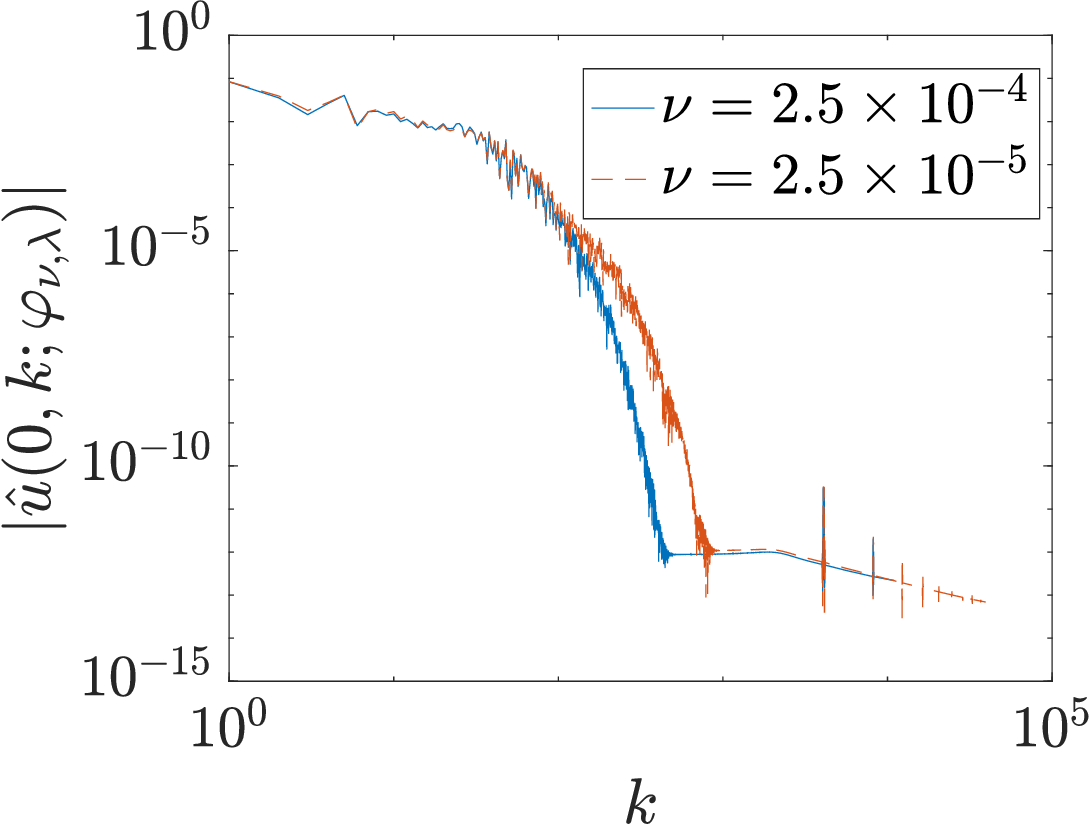}}
&{\includegraphics[scale=0.22]{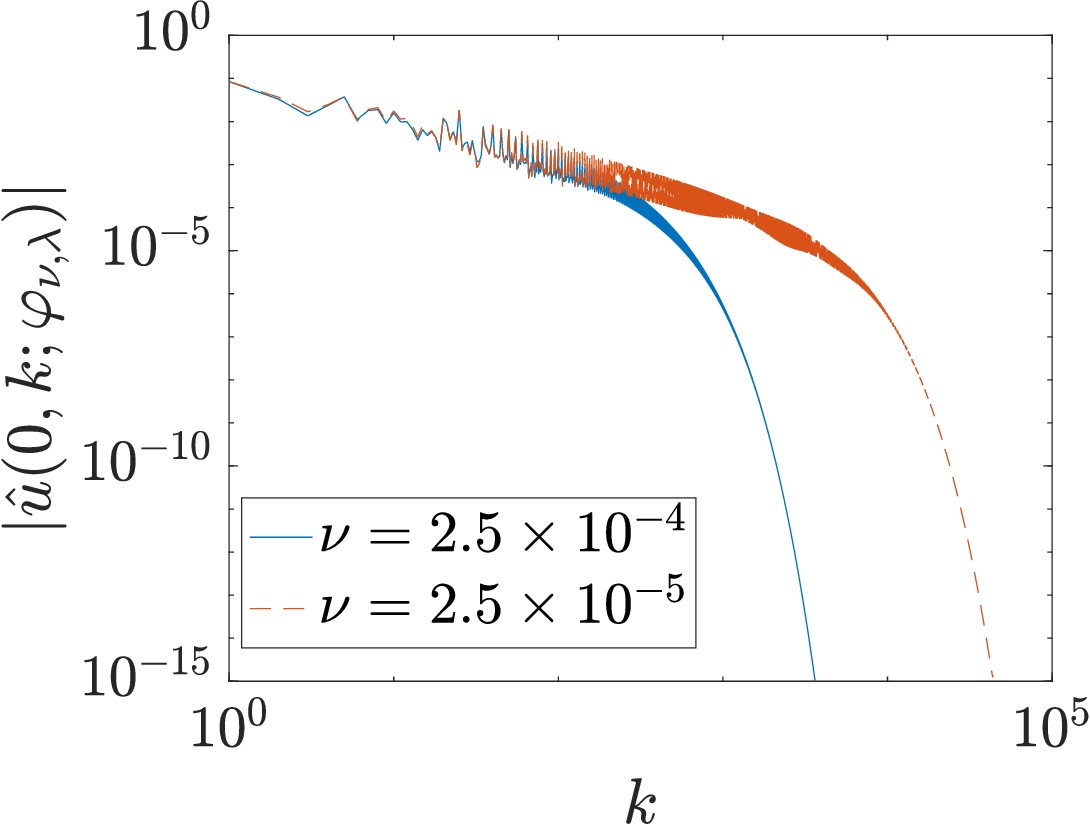}} \\
		\hline
	\end{tabular}
    \caption{Summary information about the inertial solutions found by solving \Cref{pb:1} with $\lambda =2$ (top row) and $\lambda = 5$ (bottom row), and different values of $\nu$ indicated in the individual legends. The first and the second columns represent the optimal initial conditions $\optphi$ and the final states $u(T,x;\optphi)$, with the corresponding spectra given in the third and fourth columns.}
\label{tbl:ViscComp}
\end{table}

\begin{figure}\centering
  \subfigure[]
  {
    \includegraphics[scale=0.3]{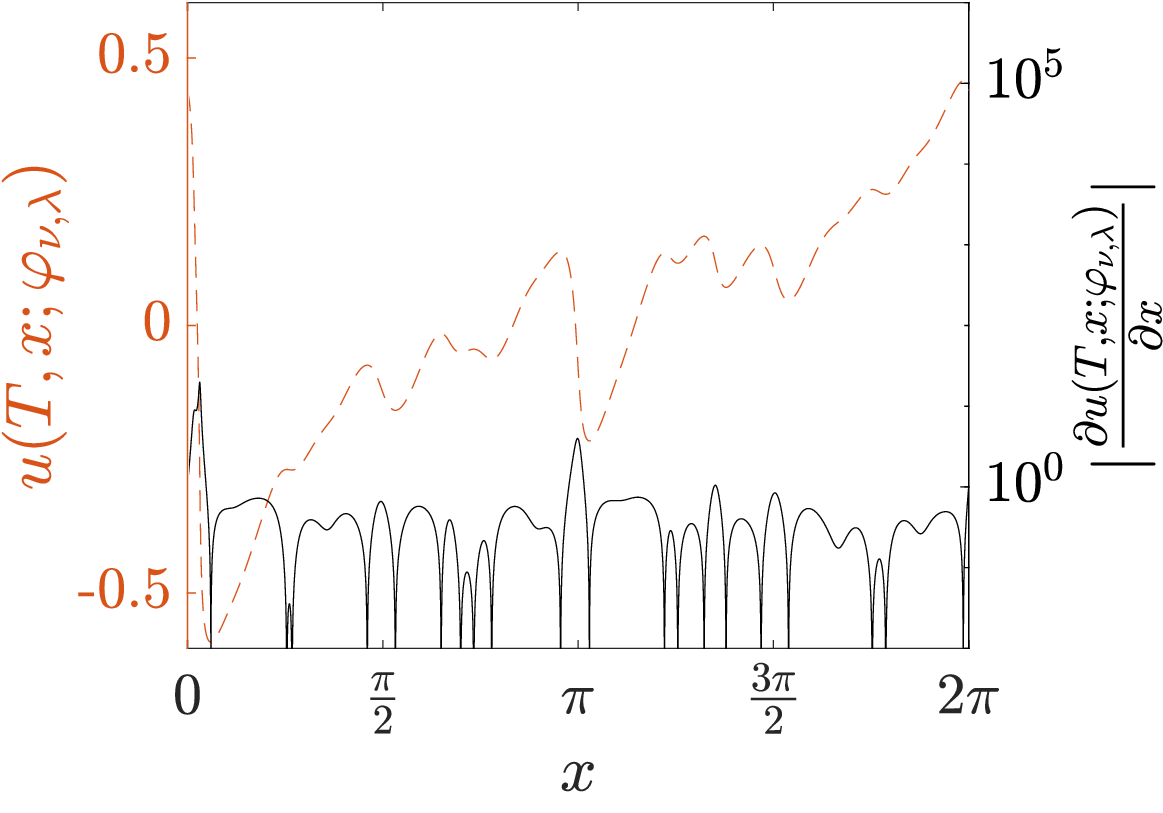}
    \label{fig:dudx2}
  }
  \subfigure[]
  {
    \includegraphics[scale=0.3]{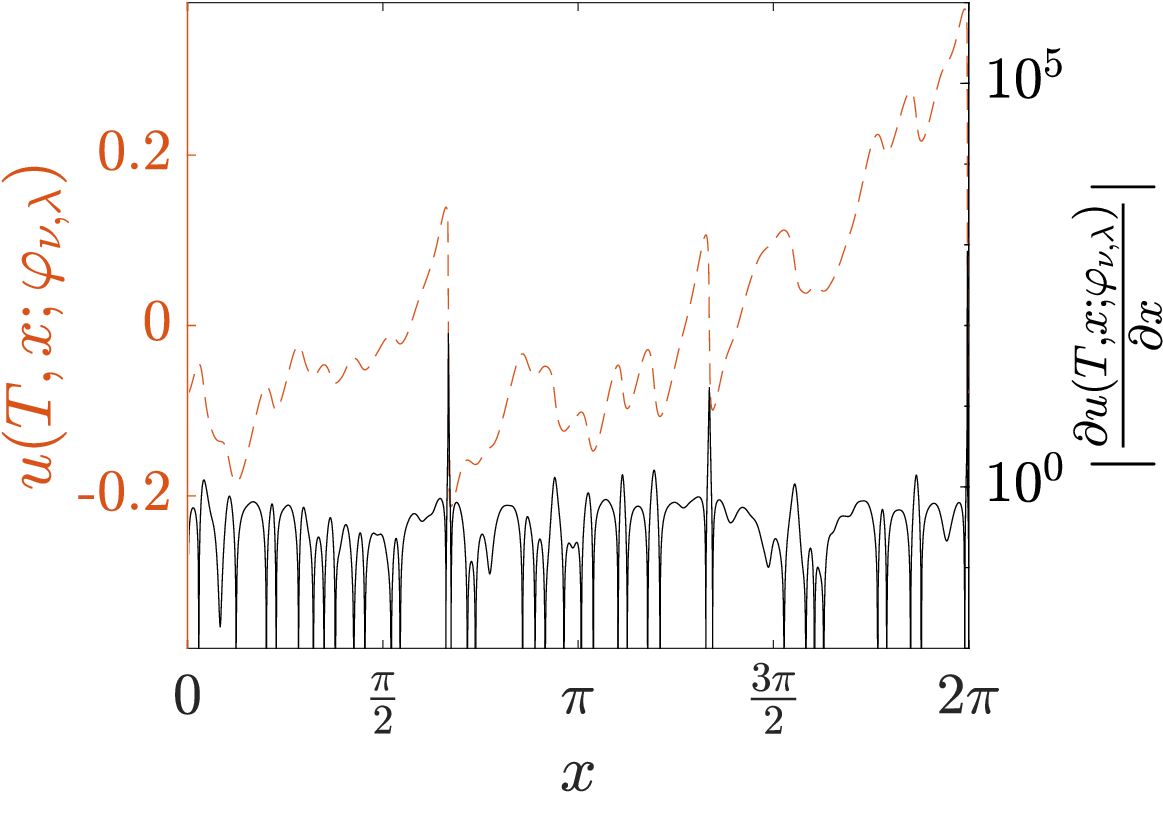}
    \label{fig:dudx3}
  }
  \subfigure[]
  {
    \includegraphics[scale=0.3]{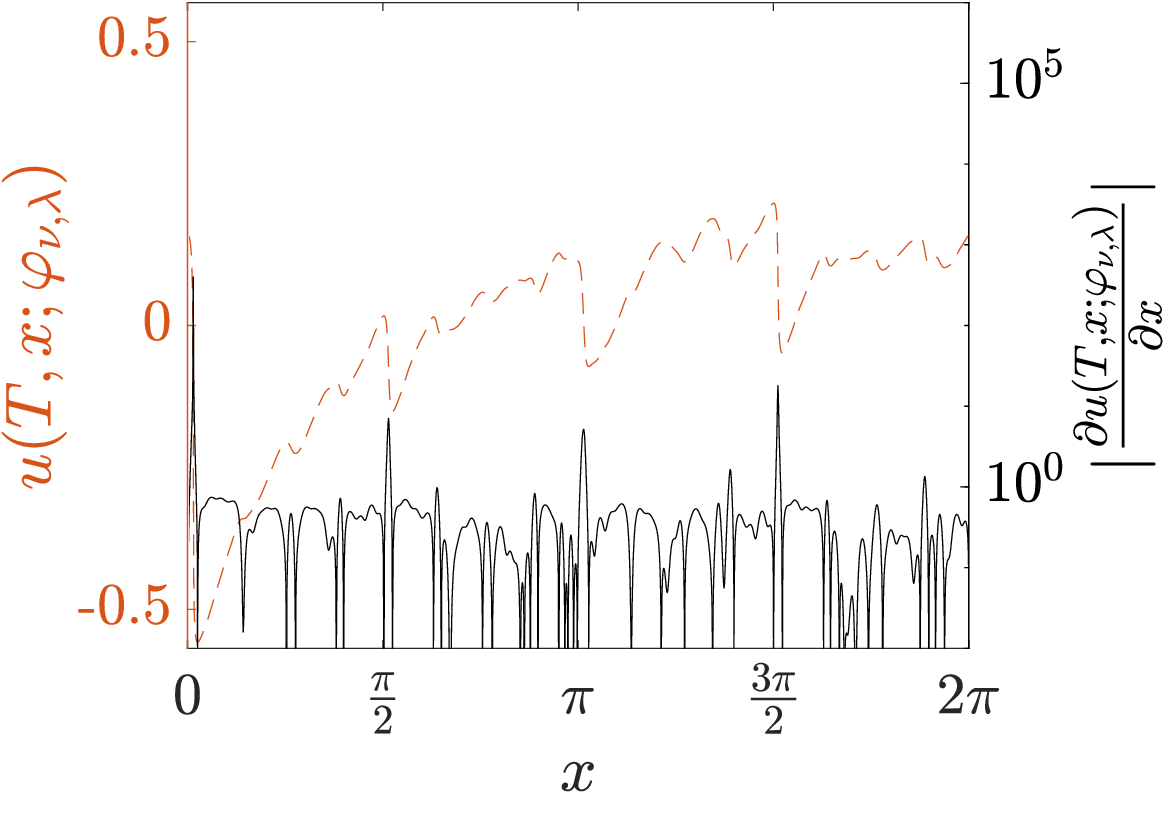}
    \label{fig:dudx4}
  }
  \subfigure[]
  {
    \includegraphics[scale=0.3]{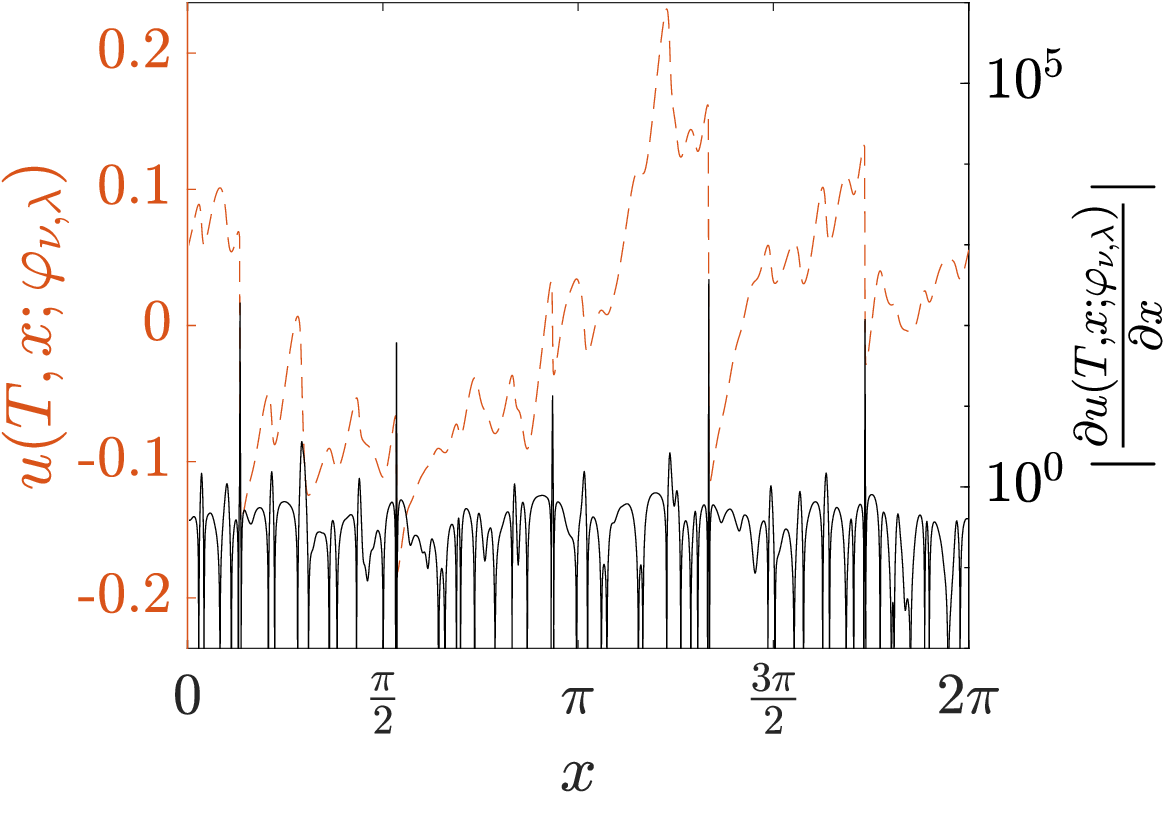}
    \label{fig:dudx5}
  }
  \subfigure[]
  {
    \includegraphics[scale=0.3]{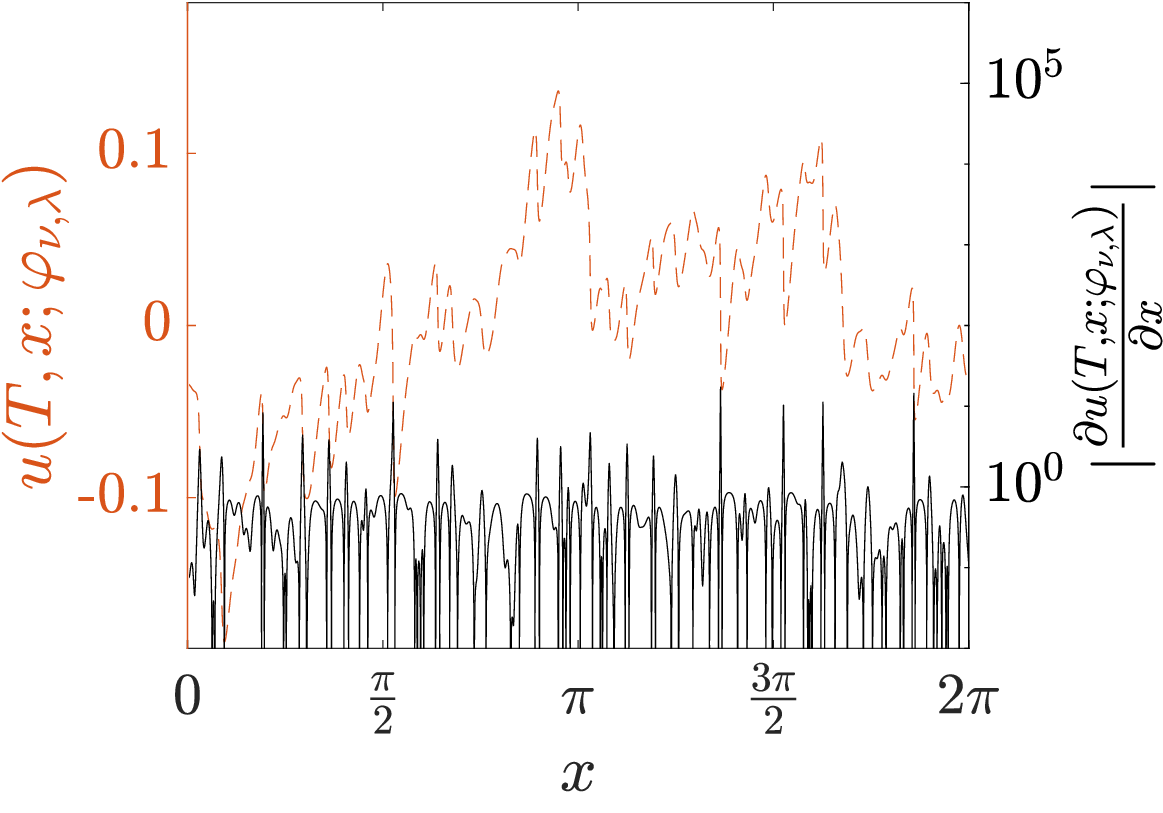}
    \label{fig:dudx6}
  }
  \subfigure[]
  {
    \includegraphics[scale=0.3]{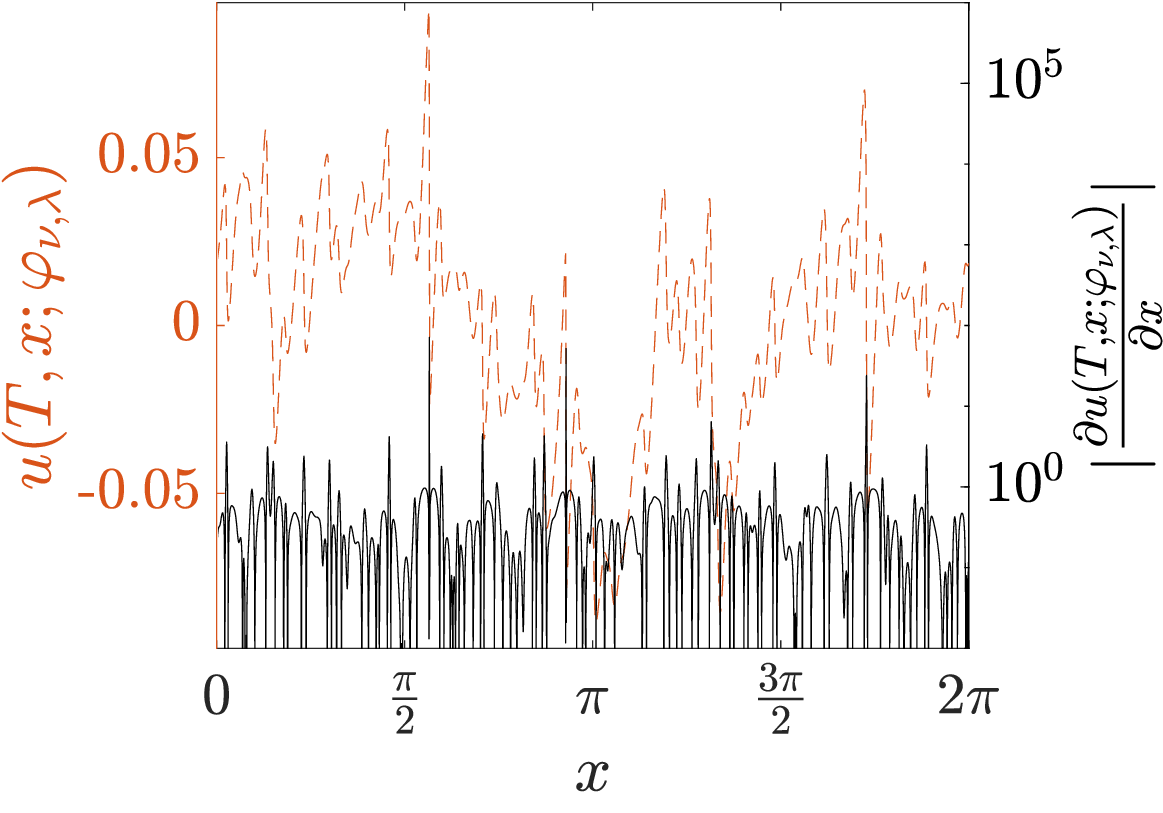}
    \label{fig:dudx7}
  }
  \caption{Solutions at the final time $u(T,x;\optphi)$ (red, left axis) and their gradients $\left|\partial u(T, x;\optphi) / \partial x \right|$ (black, right axis)) obtained by solving \Cref{pb:1} with $\nu = 2.5 \times 10^{-5}$ and (a) $\lambda = 2$, (b) $\lambda = 3$, (c) $\lambda = 4$, (d) $\lambda = 5$, (e) $\lambda = 6$, and (f) $\lambda = 7$.
}
\label{fig:dudxT}
\end{figure}

}\fi



\clearpage
\section{Conclusions\label{sec:conclusions}}

This study was motivated by an open question in turbulence research concerning the nature of fluid motions that can give rise to a self-similar energy cascade as predicted by Kolmogorov's statistical theory of turbulence \citep{Kolmogorov1941,Frisch1995book}. While the real problem concerns flows in 3D, here we consider a simple toy model, namely, the 1D viscous Burgers equation, and provide a proof of concept for a novel approach to this problem, where solutions with a {self-similar structure consistent with \Cref{def:uself}} are constructed systematically. This is done by solving a family of PDE optimization problems, cf.~\Cref{pb:1}, where we seek initial conditions for system \eqref{eq:Burgers} such that deviations from the self-similar behavior are minimized by the corresponding flow evolution over a period of time approximately equal to one eddy turnover time \eqref{eq:te}. This somewhat nonstandard PDE optimization problem is solved numerically for different values of $\nu$ and $\lambda$ using a state-of-the-art adjoint-based gradient method described in \Cref{sec:approach}. The optimal solutions $\optphi$ found in this way correspond to very small values of the objective functional \eqref{eq:J}, cf.~\Cref{fig:Jdecrease}a, indicating that they satisfy \Cref{def:uself} of self-similarity down to numerical approximation and round-off errors. \revt{Moreover, since these values approach zero (i.e., the global minimum) as the numerical resolution is refined, cf.~Figure~\ref{fig:Jrefine}, this suggests that these optimizers are in fact global.}

The solutions found fall into two distinct categories referred to as viscous and inertial distinguished by the behavior of the enstrophy $\E(u(t,\cdot;\optphi))$ which, respectively, decays and grows in these types of solutions, cf.~Figure \ref{fig:ViscInComp}h. The former solutions are uninteresting as they consist of high-wavenumber oscillations in the viscous subrange of the Fourier spectrum and as such are quickly attenuated by viscous dissipation without producing any significant energy transfer across scales. Therefore, they resemble the trivial solution $\phi_0 \equiv 0$ of the corresponding unconstrained problem, with the high-wavenumber oscillation added to satisfy the constraint $\E(\phi) = \E_0$. The fact that such physically uninteresting, yet mathematically consistent, solutions have been found demonstrates the robustness of algorithm \eqref{eq:desc}.

On the other hand, the inertial solutions lead to flow evolutions involving a self-similar, in the sense of \Cref{def:uself}, transfer of energy in the Fourier space, which in the physical space is manifested by a uniform steepening of the wave fronts, cf.~\Cref{tbl:LambComp} and \Cref{tbl:ViscComp}. As expected, optimal solutions involving such energy transfer over a large distance in the Fourier space can only be found provided the viscosity $\nu$ is sufficiently small (or, equivalently, the ``Reynolds number" is sufficiently large). Interestingly, the structure of these self-similar solutions is preserved as $\nu \rightarrow 0$, cf.~\Cref{tbl:ViscComp}, suggesting that it might persist in the limiting inviscid problem (for times before the inviscid solutions become singular). \revt{We add that front steepening in viscous Burgers flows, resulting from the competition between nonlinear amplification and viscous attenuation, is a well-studied and well-understood phenomenon \citep{FournierFrisch1983,ap11a,p12,p12b,AlbrittonDeNitti2023}. While it occurs generically in such flows, our main contribution is to demonstrate that with a suitably chosen initial data, it can take place in a self-similar manner, in the sense of Definition \ref{def:uself}.}

To the best of our knowledge, the results reported here represent the first successful effort to construct time-dependent solutions of a hydrodynamic model characterized by self-similar energy {transfer}. Both the viscous and inviscid Burgers equations are in principle analytically solvable \citep{kl04} and it is an interesting open question whether this fact could be used to rigorously justify the results presented here. \revtt{Another interesting extension is to consider forced Burgers flows where notions of self-similarity other than the one given in Definition \ref{def:uself} would be applicable. In such case, the corresponding objective functional would be minimized with respect to some properties of the forcing term added on the RHS in \eqref{eq:burg_eqn}.} While in this study we focused on a toy problem, the results obtained show promise for this approach to be \revtt{applicable to more realistic flow models, including shell models of turbulence \citep{Ditlevsen2010} and two-dimensional Navier-Stokes flows. Ultimately, the goal is to search for self-similar behavior in 3D Navier-Stokes flows which can provide new insights about hydrodynamic turbulence.}

\appendix

\section{Validation of Gradient \eqref{eq:gradL2}} \label{sec:kappa}

In order to validate the computation of gradient \cref{eq:gradL2}, which is obtained by solving the adjoint system \eqref{eq:BurgAdj} with the rather nonstandard terminal condition \eqref{eq:W} using the numerical approach described in Section \ref{sec:numerical}, we compute the G\^ateaux (directional) differential \cref{eq:dJl} in two ways: in terms of an approximation using a first-order forward finite difference formula with the step size $\epsilon$ and in terms of the  Riesz formula \cref{eq:Riesz} involving the gradient ${\grad^{L^2}} {\Jl}(\phi)$. {The} ratio of these two quantities is thus defined as
\begin{align}\label{eq:kappa}
\kappa(\epsilon) & := \frac{\epsilon^{-1} \left[ {\Jl}({\phi} + \epsilon \phi') - {\Jl}({\phi}) \right]}{\Big\langle {\grad^{L^2}} {\Jl}(\phi), \phi' \Big\rangle_{L^2(\Omega)}}, \qquad \epsilon > 0, 
\end{align}
where $\phi,\phi' \in L^2(\Omega)$ represent, respectively, the ``point" where the G\^ateaux differential is computed and the direction. Given the equivalence of the Riesz representations involving the $L^2$ and $H^1$ inner products, cf.~\eqref{eq:Riesz}, for simplicity we choose to use the former in the denominator of \eqref{eq:kappa}.

For intermediate values of $\epsilon$ the dominant errors come from the discretization of systems \eqref{eq:Burgers} and \eqref{eq:BurgAdj}, and are independent of $\epsilon$; we thus expect that $\kappa(\epsilon) \rightarrow 1$ as the discretization is refined, i.e., as $N \rightarrow \infty$ and $\Delta t \rightarrow 0$. On the other hand $\kappa(\epsilon)$ will deviate from 1 for both small and large $\epsilon$ due to, respectively, the subtractive cancellation (round-off) errors and the truncation errors in the finite-difference formula, both of which are well known effects. A detailed discussion of the interplay between these different types of errors is given in \citep[Figure 2]{Matharu_JCP517_2024}. 

To fix attention, in the tests reported here, we set $T = 0.6$, $\nu = 2.5 \times 10^{-3}$ and $\lambda=2,7$, and use the initial condition {$\phi(x) = -\sin(x)$}. Since integration in time is usually the main source of errors when employing pseudospectral methods in space, \Cref{fig:Kap} shows that indeed $\kappa(\epsilon) \to 1$ for intermediate $\epsilon \in [10^{-10},10^{-5}]$ as the temporal discretrization $\dt$ is refined with a fixed spatial resolution of $N = 16,384$ and two different perturbations $\phi'$. Similar trends were also observed in tests performed with other values of $T$, $\nu$ and $\lambda$ (not shown here for brevity).  This thus demonstrates the consistency of the computation of the gradient ${\grad^{L^2}} {\Jl}(\phi)$, both in terms of the derivation of the adjoint system \eqref{eq:BurgAdj}--\eqref{eq:W} and the numerical implementation.

\ifplots{
\begin{figure}[b]\centering
  \subfigure[]
  {
    \includegraphics[scale=0.4]{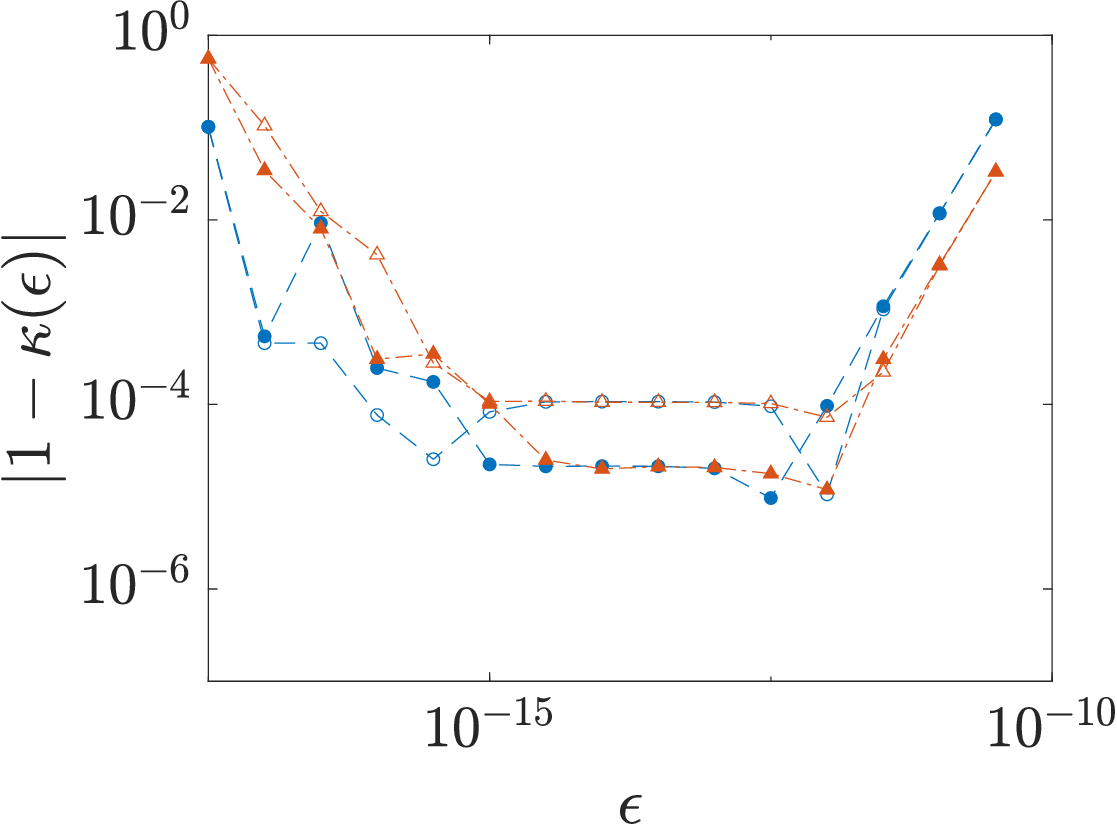}
    \label{fig:Kap2}
  }\quad
  \subfigure[]
  {
    \includegraphics[scale=0.4]{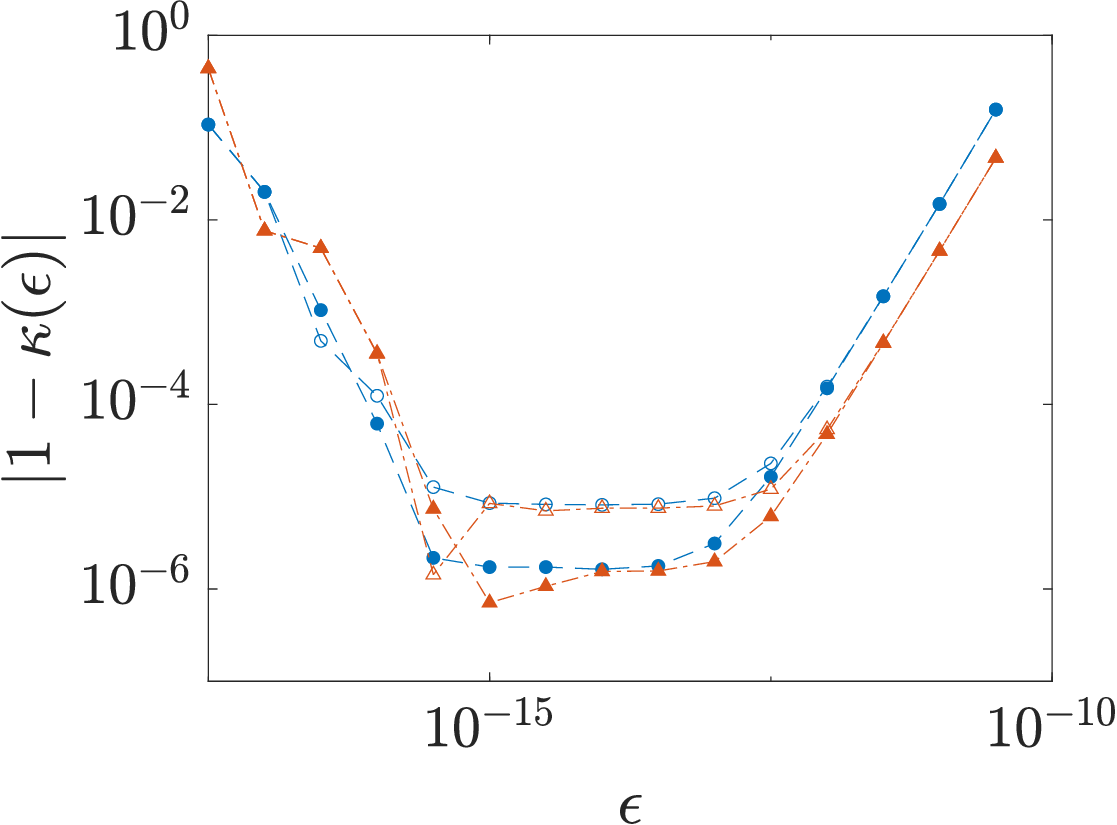}
    \label{fig:Kap7}
  }
  \caption{Dependence of $|\kappa(\epsilon) - 1|$ on $\epsilon$ for (a) $\lambda = 2$ and (b) $\lambda = 7$. Expression \eqref{eq:kappa} was evaluated using the temporal discretizations (empty symbols) $\dt = 1 \times 10^{-3}$ and (filled symbols) $\dt = 2 \times 10^{-4}$, and the perturbations (blue circles) $\phi' = \sin(-x)$ and (red triangles) $\phi' = \frac{1}{2+\sin(x)}$.}
  \label{fig:Kap}
\end{figure}
\FloatBarrier
}\fi
\section*{Acknowledgements}
B.P.~was partially supported through an NSERC (Canada) Discovery Grant RGPIN-2020-05710.
Research of T.Y.~was partly supported by the JSPS Grants-in-Aid for Scientific Research under Grant No. 24H00186.



\bibliographystyle{plainnat}
\bibliography{Bib/OptClosures,Bib/all,Bib/allPROTAS,Bib/bib_vortex_ring,Bib/control,Bib/EnergTrans,Bib/maxdEdt_biblio,Bib/blowup}
\end{document}